\documentclass[a4paper]{article}
\usepackage{amsmath}
\usepackage[english]{babel}
\usepackage{latexsym}
\usepackage{amssymb}
\usepackage{amscd}
\usepackage{amsgen,amstext,amsbsy,amsopn}

\newcommand{\version}{January 29, 2007}
\numberwithin{equation}{section}
\newcommand{\bdm}{\begin{displaymath}}
\newcommand{\edm}{\end{displaymath}}
\newcommand{\bdn}{\begin{eqnarray}}
\newcommand{\edn}{\end{eqnarray}}
\newcommand{\bay}{\begin{array}{c}}
\newcommand{\eay}{\end{array}}
\newcommand{\ben}{\begin{enumerate}}
\newcommand{\een}{\end{enumerate}}
\newcommand{\beq}{\begin{equation}}
\newcommand{\eeq}{\end{equation}}

\newcommand{\gpfR}{\mathcal{E}_R^{\mathrm{GP}}}
\newcommand{\gpdR}{\mathcal{D}_R^{\mathrm{GP}}}
\newcommand{\gpf}{\mathcal{E}^{\mathrm{GP}}}
\newcommand{\gpd}{\mathcal{D}^{\mathrm{GP}}}
\newcommand{\gpe}{E^{\mathrm{GP}}_{\varepsilon}}
\newcommand{\gpm}{\Psi_{\varepsilon}^{\mathrm{GP}}}
\newcommand{\tff}{\mathcal{E}^{\mathrm{TF}}}
\newcommand{\tfd}{\mathcal{D}^{\mathrm{TF}}}
\newcommand{\tfe}{E^{\mathrm{TF}}}
\newcommand{\tfm}{\rho^{\mathrm{TF}}}
\newcommand{\rtf}{R_0}
\newcommand{\dtf}{\mathcal{D}_0}
\newcommand{\dtfa}{\mathcal{D}_{\varepsilon}}
\newcommand{\detf}{\mathcal{T}_{\varepsilon}}
\newcommand{\br}{\mathcal{B}_R}
\newcommand{\bo}{\mathcal{B}_1}
\newcommand{\bi}{\mathcal{B}_{\varepsilon}^i}
\newcommand{\bc}{\mathcal{B}_{\frac{\spac}{2}}}

\newcommand{\trial}{\tilde{\Psi}}
\newcommand{\cut}{\chi_{\varepsilon}}
\newcommand{\latt}{\mathcal{L}}
\newcommand{\spac}{\ell_{\varepsilon}}
\newcommand{\cell}{\mathcal{Q}_{\varepsilon}}
\newcommand{\celli}{\mathcal{Q}^i_{\varepsilon}}
\newcommand{\cellj}{\mathcal{Q}^j_{\varepsilon}}
\newcommand{\chem}{\mu_{\varepsilon}}
\newcommand{\tffa}{\mathcal{E}_{\varepsilon}^{\mathrm{TF}}}

\newcommand{\tfea}{E_{\varepsilon}^{\mathrm{TF}}}
\newcommand{\tfma}{\rho_{\varepsilon}^{\mathrm{TF}}}
\newcommand{\rtfa}{R_{\varepsilon}}
\newcommand{\magnp}{\vec{A}_{\varepsilon}}
\newtheorem{lem}{Lemma}[section]
\newtheorem{teo}{Theorem}[section]
\newtheorem{pro}{Proposition}[section]

\newtheorem{cor}{Corollary}[section]
\newtheorem{rem}{Remark}[section]
\newenvironment{proof1}[1]{\mbox{} \newline \textbf{Proof of #1} \newline}{\begin{flushright} $ \Box $ \end{flushright}}
\newenvironment{proof}{\emph{Proof:}}{\begin{flushright} $ \Box $ \end{flushright}}

\begin{document}

\markboth{\scriptsize{CDY \version}}{\scriptsize{CDY \version}}

\title{Rapidly Rotating Bose-Einstein Condensates\\ in Strongly Anharmonic Traps}
\author{\hspace{-.2 cm} M. Correggi${}^{a}$, T. Rindler-Daller${}^{b}$,\\ J. Yngvason${}^{a, b}$\\
\normalsize\it \hspace{-.5 cm}\\
\hspace{-.5 cm}\normalsize\it ${}^{a}$ Erwin Schr{\"o}dinger
Institute for Mathematical Physics,\\ \normalsize\it Boltzmanngasse 9,
1090 Vienna, Austria\\ ${}^{b}$\normalsize\it Fakult\"at f\"ur Physik, Universit{\"a}t Wien,\\ \normalsize\it
Boltzmanngasse 5, 1090 Vienna, Austria}
\date{29th January 2007}

\maketitle

\begin{abstract}
We study a rotating Bose-Einstein Condensate in a strongly anharmonic
trap (flat trap with a finite radius) in the framework of 2D
Gross-Pitaevskii theory. We write the coupling constant for the
interactions between the gas atoms as $1/\varepsilon^2$ and we are
interested in the limit $\varepsilon\to 0$ (TF limit) with the angular
velocity $\Omega$ depending on $\varepsilon$. We derive rigorously the
leading asymptotics of the ground state energy and the density profile
when $\Omega$ tends to infinity as a power of $1/\varepsilon$.  If
$\Omega(\varepsilon)=\Omega_0/\varepsilon$ a ``hole'' (i.e., a region
where the density becomes exponentially small as
$1/\varepsilon\to\infty$) develops for $\Omega_0$ above a certain
critical value. If $\Omega(\varepsilon)\gg 1/\varepsilon$ the hole
essentially exhausts the container and a ``giant vortex'' develops
with the density concentrated in a thin layer at the boundary. While
we do not analyse the detailed vortex structure we prove that
rotational symmetry is broken in the ground state for ${\rm
const.}|\log\varepsilon|<\Omega(\varepsilon)\lesssim \mathrm{const.}/\varepsilon$.

\vspace{0,5cm}

\end{abstract}

\section{Introduction}

In recent years much effort, both experimental and theoretical, has
been put into the study of vortices in rotating Bose-Einstein
condensates, see, e.g., \cite{Castin}, the review \cite{Fetter} and
the monograph \cite{Afta} where extensive lists of references can be
found. Most of the theoretical research is carried out in the
framework of Gross-Pitaevskii theory, whose status as an approximation
of the quantum mechanical many-body problem was established in
\cite{Lieb2} for the non-rotating case and in \cite{Lieb1} for
rotating systems.  On the mathematical physics side an important topic
has been the vortex structure in the strong coupling (Thomas-Fermi,
TF) regime in harmonic traps when the rotational velocity is scaled
with the coupling in such a way that the number of vortices remains
finite \cite{AftaDu, Igna1, Igna2}. General results on symmetry
breaking for sufficiently large interactions or rotational velocities
and in traps of arbitrary shape were proved in the papers \cite{Seir1,
Seir2} that are not limited to the TF regime.

Recently, attention has focused on rapidly rotating condensates where
the number of vortices is much larger than unity (see, e.g.,
\cite{ECHSC,SCEMC} for experimental results).  Much of this research
has been for harmonic traps, e.g., \cite{AB, ABD, ABN1, ABN2, WBP,
WP}, where ``rapid rotation'' means a velocity close to the limiting
velocity beyond which the centrifugal forces destabilize the
condensate, but anharmonic traps (mostly quartic plus harmonic) have
also been discussed \cite{Afta1, B, baym, Fe, FiB, FZ, KB, KF,
KTU}. For harmonic traps the eigenstates of a noninteracting rotating
gas fall into Landau levels and rapid rotation implies, also for an
interacting gas, that essentially only the lowest Landau level (LLL)
is occupied. Using this fact detailed informations about the lattice
of vortices have been obtained in \cite{ABD, ABN1, ABN2, WBP}. Some
results for harmonic traps going beyond the LLL approximation are
discussed in \cite{WBP,WP}. In an anharmonic trap a restriction to the
LLL is not adequate. The reason is that the energy gap between Landau
levels is proportional to the angular velocity while the centrifugal
energy is proportional to the angular velocity squared. In an
anharmonic trap the latter can be much larger than the former, while
in a harmonic trap close to the limiting angular velocity the
potential and centrifugal energies almost cancel each other and the
LLL energy is the dominating contribution.

In the present paper we study a rapidly rotating gas in a trap that is
as far from being harmonic as possible: The gas is confined within a
finite radius $R$ and the trap is ``flat'', i.e., the confining
potential is constant (zero) inside the trap. Formally this trapping
potential can be regarded as a limit of a homogeneous potential
$V(r)\sim(r/R)^s$ with $s\to\infty$.  Such a limit naturally leads to
Dirichlet conditions at the boundary, but it is mathematically
somewhat simpler to consider the case of Neumann (or free) boundary
conditions and this is what we shall do.  In this way the interplay
between rotational effects and the nonlinear interaction terms are
brought out in a particularly clean way. Dirichlet boundary conditions
lead, in fact, to exactly the same results in the TF limit as we shall
also show.  Generalizations to homogeneous potentials with $s<\infty$
are in principle straightforward but the case $s=\infty$ merits a
special treatment because it brings out clearly the essential
differences between harmonic and anharmonic traps and also because of
some special features with respect to the breaking of rotational
symmetry.  This will be discussed in Section 2.1.

Our main results concern the density profile and the ground state
energy in the asymptotic limit when the coupling constant
$1/\varepsilon^2$ (see below) tends to infinity (TF limit) and the
rotational velocity $\Omega(\varepsilon)$ is at the same time scaled
with $\varepsilon$. (The TF limit of the 2D GP functional without
rotation is discussed in \cite{Lieb2}.) Our estimates are not sharp
enough to rigorously uncover the vortex structure of the condensate,
but the variational functions that we use and which give the correct
energy to leading order in $\varepsilon$ provide important hints about
this structure. In particular, the regimes $\Omega(\varepsilon)\sim
1/\varepsilon$ and $\Omega(\varepsilon)\gg 1/\varepsilon$ require
different variational functions, the former with a lattice of vortices
distributed over the trap and the latter with a ``giant vortex'' in
the region where the density is exponentially small.

When considering the TF limit there is an important difference between
traps that confine the gas strictly to a bounded region and traps
where the gas can spread out indefinitely. If one considers for
instance a trap given by an homogeneous potential \( V(r) = r^s \),
for some \( 0 < s < \infty \) and performs the TF limit in a naive
way, the result is trivial, namely the minimizer goes to zero and the
energy to infinity. In order to obtain a non-trivial limit it is
necessary to rescale all lengths by \( \varepsilon^{\frac{4}{2+s}}
\). In the case of infinitely high walls considered here the
characteristic length of the problem is fixed from the outset and
therefore no rescaling is needed in the TF limit. Consequences of this
difference for the question of symmetry breaking in the rotating case
will be discussed in Section 2.1.

Rapidly rotating condensates in a flat trap have been previously
studied by Fischer and Baym in \cite{FiB} and the paper of these
authors triggered, in fact, the present investigation. Our analysis
underpins and extends their general picture by rigorous estimates. We
do not, however, confirm that the transition to the ``giant vortex''
state takes place for
$\Omega(\varepsilon)\sim1/(\varepsilon^2|\log\varepsilon|)$ as implied
by Eq. (20) in \cite{FiB}. Our conclusion is rather that such a state
emerges asymptotically at all rotational velocities
$\Omega(\varepsilon)\gg 1/\varepsilon$. The reasons for this
difference are discussed in Section 2.4.

We now define the setting more precisely. The starting point is the 2D Gross-Pitaevskii (GP) energy functional
\beq
	\label{gpfR}
	\gpfR[\Psi] \equiv \int_{\br} d\vec{r} \: \left\{ |\nabla \Psi|^2 - \Omega(\varepsilon) \Psi^* L \Psi + \frac{|\Psi|^4}{\varepsilon^2}  \right\}
\eeq
where \( \br \) denotes a ball (disc) of radius \( R \) centered at the origin, \( L \) the third component of the angular momentum (i.e. \( L = -i \partial / \partial \vartheta \) in polar coordinates \( (r, \vartheta) \)), \( \Omega(\varepsilon) \) the angular velocity and \( \varepsilon \) is a nonnegative, small parameter.
\newline
The functional is defined on the domain\footnote{By Sobolev immersion \( H^1(\br) \) is contained in \( L^4(\br) \), so the functional is well defined on \( \gpdR \).} 
\beq
	\label{domain}
    	\gpdR = H^1(\br). 
\eeq
We also define
\beq
    	\label{minimization}
    	\gpe (R) \equiv \min_{   
		\bay
			\Psi \in \gpdR \\
                	\| \Psi \|_2 = 1
         	\eay} \gpfR[\Psi]
\eeq
and denote by \( \gpm \) a corresponding minimizer\footnote{Any result
for \( \gpm \) stated in the following is meant to be true for any
minimizer, if it is not unique.}. Indeed, for any \(
\Omega(\varepsilon) < \infty \), one can prove (see for instance
\cite{Seir1}) that the functional is bounded from below and there
exists at least one minimizer.

{}From the physics point of view a minimizer of (\ref{gpfR}) describes
the macroscopic wave function (the wave function of the condensate) 
of a Bose-Einstein condensate in the
rotating reference frame. The 2D description is a simplification that
is justified either in the limit of thin (``disc shaped'') 3D traps,
or traps that are very elongated along the rotational axis (``cigar
shaped'' traps) so that the 3D wave function is essentially constant
along this axis. In both cases the coupling $1/\varepsilon^2$ is
proportional to $Na/h$ where $a$ is the scattering length of the two-body potential (for its definition see, e.g., the Appendix in \cite{Lieb3}) , $N$ the
particle number and $h$ the extension of the 3D trap along the
rotational axis.\footnote{In ``thin'' traps a different formula
applies at extreme dilution where the coupling becomes independent of
$a$ and depends logarithmically on the average density (see
\cite{SY}).}

In the non-rotating case, \( \Omega(\varepsilon) = 0 \), the minimizer is actually unique, by the strict convexity of the functional, and it is given by the (normalized) constant function,
\bdm
    \gpm\big|_{\Omega = 0} =  \frac{1}{\sqrt{\pi} R}.
\edm
The ground state energy is then
\bdm
    \gpe(R)\big|_{\Omega = 0} = \frac{1}{\pi R^2 \varepsilon^2}.
\edm
If the angular velocity is different from zero the minimizer may not be unique since a rotational symmetry breaking occurs for \( \Omega(\varepsilon) \) above a certain threshold value as will be shown in Section 2.1.

We point out that the dependence on the radius \( R \) of the trap can be scaled out:
\beq
    \label{gpe}
    \gpe \equiv \gpe (1) = R^2 \gpe (R)
\eeq
so that, without loss of generality, we can choose \( R = 1 \) and denote the functional by \( \gpf[\Psi] \).

The GP functional can be rewritten in the following form that we are going to use:
\beq
    \label{gpfmagnetic}
    \gpf[\Psi] = \int_{\bo} d\vec{r} \: \left\{ \left| \left( \nabla - i \magnp \right) \Psi \right|^2 - \frac{\Omega(\varepsilon)^2 r^2 | \Psi |^2}{4} + \frac{|\Psi|^4}{\varepsilon^2}  \right\}
\eeq
where \( \magnp \) is the vector potential associated with the rotation, i.e.,
\beq
    \label{magnp}
    \magnp(\vec{r}) \equiv \frac{\Omega(\varepsilon)}{2} \hat{z} \times \vec{r},
\eeq
with $\hat z$ the unit vector in the $z$-direction.  In
\eqref{gpfmagnetic} one can recognize an analogy with the
Ginzburg-Landau (GL) functional (see, e.g., \cite{BR}) in the theory
of superconductivity. The vector potential \eqref{magnp} is in this
context due to a uniform magnetic field, while the wave function of
the condensate is the GL order parameter (density of Cooper
pairs). Using the \( L^2-\)normalization of the minimizer, the analogy
can be made even closer, namely the minimization problem in
\eqref{minimization} is equivalent to the minimization of the
functional
\bdm
	{\gpf}^{\prime}[\Psi] = \int_{\bo} d\vec{r} \: \left\{ \left| \left( \nabla - i \magnp \right) \Psi \right|^2 - \frac{\Omega(\varepsilon)^2 r^2 | \Psi |^2}{4} + \frac{\left(1 - |\Psi|^2 \right)^2}{\varepsilon^2}  \right\}
\edm
over \(L^2\)-normalized functions. At this point, however, an
important difference becomes evident, namely the presence of the
centrifugal energy (the second term in the expression above), which in
the GL context could be interpreted as an electric field. This
contribution, usually not present in the GL functional, is
proportional to the square of the angular velocity and we are going to
see that, in the regimes we are considering, it is responsible for a
rather different behavior of the minimizer compared to GL
theory. Another important difference between the GP and the GL
minimization problems is the \( L^2\)-normalization condition, that
prevents, for instance, the minimizer from being identically zero, as
it can be in the GL case. It also gives rise to an additional term
(chemical potential) in the variational equation associated to
\eqref{minimization}.

In the next Section 2 we introduce some notations and state the main
results of this paper. We first discuss the problem of spontaneous
symmetry breaking in the ground state, then we study the regimes \(
\Omega(\varepsilon) \ll 1/\varepsilon \) (Section 2.2), \(
\Omega(\varepsilon) \sim 1/\varepsilon \) (Section 2.3) and \(
\Omega(\varepsilon) \gg 1/\varepsilon \) (Section 2.4). Section 3 is
devoted to the proofs, while in Section 4 we comment on the results
and perspectives.

\section{Main Results}

\subsection{Spontaneous Symmetry Breaking in the Ground State}

The GP functional for a rotating 2D condensate in a general trap has
already been studied in \cite{Seir1}. A very interesting phenomenon
generated by the rotation is the spontaneous breaking of rotational
symmetry in the ground state. If the trap potential is polynomially
bounded at infinity one can prove (see Theorem 4 in \cite{Seir1}) that
for any fixed angular velocity \( \Omega \), there exists \(
\varepsilon_{\Omega} \) such that, if \( \varepsilon <
\varepsilon_{\Omega} \), no ground state of the GP functional is an
eigenfunction of the angular momentum. The rotational symmetry of the
functional is then spontaneously broken at the level of the ground
state. An important consequence is that the minimizer is no longer
unique, since a rotation by an arbitrary angle gives rise to a state
with the same energy.

A crucial ingredient of the proof in \cite{Seir1} is that, in a
polynomially bounded potential trap, the density of the minimizer
tends to zero as \( \varepsilon \to 0 \). In fact, Theorem 4 in
\cite{Seir1} is not true in the case of a trap with infinitely high
walls as we are considering and we actually expect the opposite
behavior: If \( \Omega \) is kept fixed, then for \( \varepsilon \)
sufficiently small the ground state is an eigenfunction of the angular
momentum, and after an appropriate choice of a constant phase factor,
a unique, strictly positive radial function. This difference can be understood by  noting that in a trap of radius $R$ the kinetic energy of a vortex is of the order $R^{-2}|\log \varepsilon|$ for small
 $\varepsilon$. Thus, if $R$ is fixed, an angular velocity of   order \(
| \log \varepsilon | \) is needed in order to create vortices. In a polynomially bounded trap, on the other hand, the effective radius of the condensate increases as $\varepsilon\to 0$ and the critical velocity for the creation of a vortex behaves as $\varepsilon^{4/(s+2)}|\log\varepsilon|$ for a trap potential $\sim r^s$, cf. the remark at the end of Section 3 in \cite{Seir1}. Any  fixed $\Omega$ thus exceeds the critical velocity as $\varepsilon\to 0$ if $s<\infty$. 

Despite this difference,  our  proof of symmetry breaking is obtained partly by a modification of
the arguments of Theorem 2 in \cite{Seir1}.
The following
Proposition \ref{instability} states that for angular velocities smaller than \(
1/\sqrt{\pi} \varepsilon \), symmetric vortices of degree higher than 1 are
unstable\footnote{In the opposite regime of weak coupling and fixed $\Omega$, vortices of degree 2 or higher may be energetically favorable in anharmonic traps \cite{Lund1}.}.

\begin{pro}[Instability for Higher Vorticity]
	\label{instability}
	\mbox{}	\\
	Let \( \Psi_n(\vec{r}) \), \( n \geq 2 \), be the unique minimizer of \( \gpf[\Psi] \) on the subspace of functions with angular momentum \( n \), i.e., on  \( \{ \Psi \in \gpd \: | \: L \Psi = n \Psi \}\). For any \( \Omega(\varepsilon) \leq 1/\sqrt{\pi} \varepsilon \), \( \Psi_n \) is unstable, i.e.,  it is not a local minimizer of \( \gpf[\Psi] \).
\end{pro}

\begin{proof}
	From the variational equation satisfied by the radial part of \( \Psi_n \equiv \xi_n(r) e^{in\vartheta} \),
    	\bdm
        	- \Delta \xi_n + \frac{n^2 \xi_n}{r^2} - \Omega(\varepsilon) n \xi_n + \frac{2 \xi_n^3}{\varepsilon^2} = \mu_{n}(\varepsilon) \xi_n
   	\edm
    	where the chemical potential \( \mu_n(\varepsilon) \) is fixed by the \(L^2\)-normalization of \( \Psi_n \), it is not hard to prove by a rearrangement argument (see, e.g., Lemma 1 in \cite{Seir1}) that $ \xi_n $ is a positive non-decreasing function, $ \xi_n(r) = O(r^n) $ as $ r \to 0 $ and $ \xi^{\prime}_n(1) = 0 $, i.e., $ \xi_n $ satisfies Neumann boundary conditions. Moreover by a subharmonicity argument\footnote{For a similar proof see, e.g., Lemma 2.1 in \cite{Lieb2}.} we can also prove the bound: 
    	\beq	
	\label{Linftybound}
       		\left\| \Psi_n \right\|^2_{L^{\infty}(\bo)} \leq \frac{\varepsilon^2}{2} \left\{ \mu_n(\varepsilon) + \Omega(\varepsilon) n - n^2 \right\}
    	\eeq
    	Indeed, suppose that $ n \geq 1 $ and the opposite is true, then setting
	\bdm
		\mathcal{B}^{>} \equiv \left\{ r \in (0,1) \: \Big| \: \xi_n^2(r) > \varepsilon^2 \left( \mu_n(\varepsilon) + \Omega(\varepsilon) n - n^2 \right)/2 \right\} 
    	\edm
	we can have two possibilities: Either $ \mathcal{B}^{>} = \emptyset $, and then the result easily follows, or it is an open interval, $ \mathcal{B}^{>} \equiv (R^{>},1) $, by monotonicity of $ \xi_n $, and $ \Delta \xi_n|_{\mathcal{B}^{>}} > 0 $. In this case, by integrating $ \Delta \xi_n $ over $ \mathcal{B}^{>} $ and using Neuman boundary conditions, one has
	\bdm
		\int_{\mathcal{B}^{>}} \Delta \xi_n r dr= - R^{>} \xi^{\prime}_n(R^{>}) > 0  
	\edm
	which is a contradiction because $ \xi_n $ in non-decreasing.
	\newline
	The rest of the proof coincides with the proof of Theorem 2 in \cite{Seir1}. Using the estimate \eqref{Linftybound}, we can extract, as in (2.33) in \cite{Seir1}, a sufficient condition on the chemical potential for instability of the corresponding vortex: The symmetric vortex of degree\footnote{The variational parameter $ d \in \mathbb{N} $ is involved in the definition of a suitable trial function used in Theorem 2 in \cite{Seir1}. The requirement $ n \geq d $ is necessary, otherwise such a trial function does not belong to $ H^1(\bo) $.} \( n \geq d \in \mathbb{N} \) is unstable if
	\bdm
		- (d-1)^2 \chem + (d^2 -1) \Omega(\varepsilon) n - (d-1)^2 n^2 < 0
	\edm
	or, choosing $ d = 2 $,
    	\bdm
        	- \mu_{n}(\varepsilon) + 3 \Omega(\varepsilon) n - n^2 < 0.
    	\edm
    	From the definition of the chemical potential and Schwarz's inequality it also follows that
    	\bdm
        	\mu_n(\varepsilon) = \int_{\bo} d\vec{r} \: \left\{ \left( \nabla \xi_n \right)^2 + \frac{n^2 \xi_n^2}{r^2} - \Omega(\varepsilon) n \xi_n^2 + \frac{2 \xi_n^4}{\varepsilon^2} \right\} \geq n^2 - \Omega(\varepsilon) n + \frac{2}{\pi \varepsilon^2}.
    	\edm
    	Inserting this bound in the condition above, we have instability if
    	\bdm
        	n^2 - 2 \Omega(\varepsilon) n + \frac{1}{\pi \varepsilon^2} > 0.
    	\edm
    	Hence any vortex of order \( n \geq 2 \) is unstable, provided  \( \Omega(\varepsilon) \leq 1/\sqrt{\pi} \varepsilon \).
\end{proof}

{}From Prop.\  \ref{instability}  it follows that in order to prove the symmetry breaking in the ground state 
for  a given $ \Omega(\varepsilon) \leq 1/\sqrt{\pi} \varepsilon $ it is sufficient to show that a rotationally symmetric vortex of degree smaller or equal to 1 
cannot be a minimizer of the GP functional at this angular velocity.
This in turn can be achieved by exploiting some energy estimates. Let us first define 
\bdm 
	E_n(\varepsilon) \equiv \min_{\| \xi \|_{L^2({\mathcal B}_1)} = 1} \int_{\bo} d\vec{r} \: \left\{ \left| \nabla \xi \right|^2 + \frac{n^2 \xi^2}{r^2} + \frac{\xi^2}{\varepsilon^2} \right\}
\edm
and
\bdm
    \Omega_n(\varepsilon) \equiv E_{n+1}(\varepsilon) - E_n(\varepsilon)
\edm
so for  \( \Omega(\varepsilon) > \Omega_{\bar{n}}(\varepsilon) \) no symmetric vortex of degree \( n \leq \bar{n} \) can be a global minimizer of the GP functional.  Since \( \Omega_n(\varepsilon) \leq (2n+1) \Omega_0(\varepsilon) \) (see Lemma 3 in \cite{Seir1}) we can use an upper bound on $E_{1}(\varepsilon) - E_0(\varepsilon)$ to prove the symmetry breaking.

Proposition \ref{instability} applies only for $ \Omega(\varepsilon) \leq 1/\sqrt{\pi} \varepsilon $ but symmetry breaking can, in fact, be proved  for $ \Omega(\varepsilon) \lesssim C/\varepsilon $ with an arbitrary constant $ C $ by using  Theorem \ref{energy} that is proved later in the paper. Hence a part of the proof of the next proposition will be  postponed to the end of Section 2.3.

\begin{pro}[Symmetry Breaking in the Ground State]
    	\label{breaking}
	\mbox{}	\\
	For \( \varepsilon \) sufficiently small, no minimizer of \( \gpf[\Psi] \) is an eigenfunction of the angular momentum, if
    	\bdm
        	6 |\log \varepsilon | + 3 < \Omega(\varepsilon) \lesssim \frac{C}{\varepsilon}
    	\edm	
	for any constant $ C \in \mathbb{R}^+ $.
\end{pro}

\begin{proof}
    	Using the normalized trial function
    	\bdm
        	\xi(r) \equiv c_{\varepsilon}
        	\left\{
        	\begin{array}{ll}
            		\displaystyle{\frac{r}{\varepsilon}}        &   \mbox{if} \:\:\:\: 0 \leq r \leq \varepsilon    \\
            		\mbox{}     &   \mbox{} \\
         	   	1       &   \mbox{otherwise}
        	\end{array}
        	\right.
    	\edm
    	we can prove the upper bound
    	\bdm
    		E_1(\varepsilon) \leq \frac{1}{\pi \varepsilon^2} + \log \left( \frac{1}{\varepsilon^2} \right) + 1.
    	\edm
    	Since \( \Omega_n(\varepsilon) \leq (2n+1) \Omega_0(\varepsilon) \) (see Lemma 3 in \cite{Seir1}) and \( E_0(\varepsilon) = \frac{1}{\pi \varepsilon^2} \), we get
    	\bdm
        	\Omega_1(\varepsilon) \leq 3 \left[ \log \left( \frac{1}{\varepsilon^2} \right) + 1 \right].
    	\edm
	Hence no symmetric vortex of degree \( \leq 1 \) can be a global minimizer of the GP functional if \( \Omega(\varepsilon) \geq 6 |\log \varepsilon | + 3 \). On the other hand vortices of degree higher or equal to 2 are excluded by Proposition \ref{instability} provided that \( \Omega(\varepsilon) \lesssim 1/\sqrt{\pi} \varepsilon \). The proof of symmetry breaking for general  $ \Omega(\varepsilon) \sim 1/\varepsilon $ will be given at the end of Section 2.3.
\end{proof}

\subsection{Energy and density for $\Omega(\varepsilon)\ll 1/\varepsilon$}

If \( \Omega(\varepsilon) \ll 1/\varepsilon \), the rotation has no leading order effect in the TF regime. More precisely the energy asymptotics is the same as for a non-rotating condensate, and the density profile, \( |\gpm|^2 \), converges to the normalized constant function, namely the minimizer of the GP functional without rotation:

\begin{pro}[Energy and Density Asymptotics]
	\label{norotation}
	\mbox{}	\\
	For any \( \Omega(\varepsilon) \) such that \( \lim_{\varepsilon \to 0} \varepsilon \Omega(\varepsilon) = 0 \) and for \( \varepsilon \) sufficiently small
    	\beq
        	\label{energynorot}
        	\varepsilon^2 \gpe = \frac{1}{\pi} - O(\varepsilon^2 \Omega(\varepsilon)^2)
    	\eeq
    	\beq
        	\label{convnorot}
        	\left\| |\gpm|^2 - 1/\pi \right\|_{L^1(\bo)} = O(\varepsilon \Omega(\varepsilon)).
    	\eeq
\end{pro}

\begin{proof}
	Since \( \| \gpm \|_{L^2(\bo)} = 1 \) and \( R = 1 \), we have 
	\bdm
		\int_{\bo} d\vec{r} \: r^2 |\gpm|^2 \leq 1 
	\edm
	and \( \| \gpm \|^4_{L^4(\bo)} \geq 1/\pi \) by Schwarz's inequality. Hence \eqref{gpfmagnetic} leads to the lower bound,
    	\bdm
        	\varepsilon^2 \gpe \geq \frac{1}{\pi} - \frac{\varepsilon^2 \Omega(\varepsilon)^2}{4}.
    	\edm
    	The upper bound is obtained by evaluating the functional on the trial function \( 1/\sqrt{\pi} \), namely
    	\bdm
        	\varepsilon^2 \gpe \leq \frac{1}{\pi}.
    	\edm
    	Moreover, since \( \| \gpm \|_{L^2(\bo)} = 1 \), this estimate implies 
    	\bdm
        	\left\| |\gpm|^2 - 1/\pi \right\|^2_{L^2(\bo)} \leq \frac{\varepsilon^2 \Omega(\varepsilon)^2}{4}
    	\edm
    	The \( L^1-\)convergence of the density profile now follows by Schwarz's inequality.
\end{proof}

We stress that the result above says nothing about the fine structure
of the minimizer and also nothing about its uniqueness. As far as the
density profile is concerned, the first critical velocity at which
some new effect comes into play is \( \Omega(\varepsilon) \sim
1/\varepsilon \) as it will be discussed in the next subsection.  On
the other hand, the fine structure of \( \gpm \) depends on the
angular velocity, even if \( \Omega(\varepsilon) \ll 1/\varepsilon \).
If \( \Omega(\varepsilon) \) is simply a constant and \( \varepsilon
\) is sufficiently small, it is not hard to see that the minimizer is
unique. More precisely, it is a radial function (and hence an
eigenfunction of the angular momentum) which can be chosen strictly
positive. In this case the result in \eqref{convnorot} can be improved
and the convergence can be extended to \( L^{\infty}(\bo) \).

According to the discussion in \cite{AftaDu} and the rigorous analysis in \cite{Igna1,Igna2} of rotating Bose-Einstein condensates in harmonic traps, the first critical velocity for the occurrence of vortices, i.e. isolated zeros of the minimizer, is in that case\footnote{The overall factor \( \varepsilon \) in the critical velocities \eqref{critical} is due to the scaling mentioned in Section 1 and Section 2.1.}  of the order \( \Omega(\varepsilon) \sim \varepsilon | \log \varepsilon| \). More precisely, if \( \tilde{\Omega}_d(\varepsilon) < \Omega(\varepsilon) < \tilde{\Omega}_{d+1}(\varepsilon) \), where 
\beq
\label{critical}
	\tilde{\Omega}_d(\varepsilon) \equiv c \: \varepsilon \left[ |\log \varepsilon | + (d-1)\log|\log \varepsilon| \right] ,
\eeq
the minimizer has exactly \( d \) vortices of degree 1. A similar behavior was shown in \cite{Serf1} for a slightly different model of superfluids.
\newline
Such results together with the considerations in Section 2.1 suggest that in a flat trap vortices start to occur  if \( \Omega(\varepsilon) \sim  | \log \varepsilon | \) and the rotational symmetry can be broken. The spontaneous symmetry breaking cannot be seen at the level of the density profile \( | \gpm |^2 \) however, because the average size of each vortex is very small (area of the core of order \( \varepsilon \)) in the TF limit. The total vorticity of the minimizer is proportional to the angular velocity, provided that \( \Omega(\varepsilon) \gg |\log\varepsilon| \), and therefore, as long as \( \Omega(\varepsilon) \ll 1/\varepsilon \), the region covered by the vortex cores has Lebesgue measure zero in the limit $\varepsilon \to 0$, in accord with \eqref{convnorot}.

\subsection{The Regime $\Omega(\varepsilon)\sim1/\varepsilon$}

In the regime \( \Omega(\varepsilon) \sim 1/\varepsilon \) the rotation is so fast that it modifies the density profile itself: Since the centrifugal energy in \eqref{gpfmagnetic} is of the same order of the non-linear term, it is no longer convenient for the condensate to be uniformly distributed over the trap, like in the non-rotating case. Such an effect can be seen at a macroscopic level, namely the density profile converges to a non-constant function, which minimizes a TF-like functional.

Before stating the main results we first need some new notations. Without loss of generality we can assume that $ \Omega_0 \equiv \varepsilon \Omega(\varepsilon) $ is a constant independent of $ \varepsilon $. Moreover, for any \( \Omega_0 > 0 \), we introduce the \emph{TF functional},
\beq
	\label{tff}
    	\tff[\rho] \equiv \int_{\bo} d\vec{r} \: \left\{ \rho^2 - \frac{\Omega_0^2 r^2 \rho}{4} \right\},
\eeq
defined on the domain
\beq
    	\tfd = \left\{ \rho \in L^2(\bo) \: | \: \rho \geq 0 \right\}.
\eeq
The functional above has a unique minimizer, \( \tfm \), and we denote
\beq
	\label{tfe}
    	\tfe \equiv \min_{   
		\bay
                	\rho \in \mathcal{D}^{TF} \\
                	\int \rho = 1
            	\eay} \tff[\rho] = \tff[\tfm].
\eeq
The minimizer \( \tfm \) can be explicitly calculated:
\beq
	\label{tfm}
    	\tfm(r) = \left\{
        \begin{array}{ll}
            	\displaystyle{\frac{1}{\pi}} - \frac{\Omega_0^2}{16} (1-2r^2)                   &   \mbox{if} \:\:\:\: \Omega_0 \leq \displaystyle{\frac{4}{\sqrt{\pi}}}    \\
            	\mbox{}     &   \mbox{} \\
            	\left[ \displaystyle{\frac{\Omega_0^2}{8}} (r^2-1) + \frac{\Omega_0}{2 \sqrt{\pi}} \right]_+    &   \mbox{if} \:\:\:\: \Omega_0 > \displaystyle{\frac{4}{\sqrt{\pi}}}
        \end{array}
        \right.
\eeq
where \( [ \: \cdot \: ]_+ \) stands for the positive part, and the ground state energy is
\beq
    	\tfe = \left\{
        \begin{array}{ll}
            	\displaystyle{\frac{1}{\pi}}-\frac{\Omega_0^2}{8} - \frac{\pi \Omega_0^4}{768}      &   \mbox{if} \:\:\:\: \Omega_0 \leq \displaystyle{\frac{4}{\sqrt{\pi}}}    \\
            	\mbox{}     &   \mbox{} \\
            	\displaystyle{\frac{\Omega_0}{4}} \left(\frac{8}{3 \sqrt{\pi}} - \Omega_0 \right)   &   \mbox{if} \:\:\:\: \Omega_0 > \displaystyle{\frac{4}{\sqrt{\pi}}}.
        \end{array}
        \right.
\eeq

We point out that, if \( \Omega_0 > \frac{4}{\sqrt{\pi}} \), \( \tfm \) has a ``hole'', i.e., a macroscopic region where it is identically zero, centered at the origin: With
\beq
    	\label{rtf}
    	\rtf \equiv \sqrt{1 - \frac{4}{\sqrt{\pi} \Omega_0}}
\eeq
we have \( \tfm(r) = 0 \), for any \( r \leq \rtf \). We also define
\beq
    	\label{dtf}
    	\dtf \equiv \mathrm{supp}(\tfm) = \{ \vec{r} \in \bo \: | \: r \geq \rtf \}.
\eeq

The first result concerns the energy asymptotics:

\begin{teo}[Energy Asymptotics]
	\label{energy}
	\mbox{}	\\
	For any \( \Omega_0 > 0 \) and for \( \varepsilon \) sufficiently small
	\beq\label{energyest}
        	\varepsilon^2 \: \gpe = \tfe + \mbox{O}(\varepsilon | \log\varepsilon |).
	\eeq
\end{teo}

The leading order term, proportional to \( 1/\varepsilon^2 \), in the asymptotic expansion of \( \gpe \) is due to the centrifugal bending of the profile while the remainder (of the order \( {|\log \varepsilon|}/{\varepsilon} \)) is the contribution coming from the fine structure of the minimizer. Indeed, \( \gpm \) is expected to carry a very large number (of the same order as \( \Omega(\varepsilon) \)) of vortices of degree 1. As suggested by the trial function \eqref{trial} used  in the proof of Theorem \ref{energy} (see also \cite{FiB}), such vortices should be distributed over a lattice with a spacing of order \( \sqrt{\varepsilon} \), so that the average vortex core covers an area proportional to \( \varepsilon \). A simple argument (see, for instance, \cite{Beth2}) shows that the kinetic energy of each vortex is of the order \( |\log \varepsilon| \). This explains why the total energy contribution of vortices produces a remainder of the order \( {|\log \varepsilon|}/{\varepsilon} \). 

As stated in the Introduction, the results proved in Theorem \ref{energy} and in the rest of this Section also hold in the case of Dirichlet boundary conditions (see the Remark \ref{Dirichlet} in Section 3.1). The crucial point is that the limiting functional \eqref{tff} contains no kinetic energy and hence boundary conditions become irrelevant in the TF limit, at least to the leading order.  

The convergence of the profile \( | \gpm |^2 \) to \( \tfm \) is a straightforward consequence of Theorem \ref{energy}:

\begin{cor}[Density Asymptotics]
	\label{profile}
	\mbox{}	\\
	For any \( \Omega_0 > 0 \) and for \( \varepsilon \) sufficiently small,
	\beq
        	\label{L1profile}
        	\left\| | \gpm |^2 - \tfm \right\|_{L^1(\bo)} = O(\sqrt{\varepsilon |\log \varepsilon|}).
	\eeq
\end{cor}

If \( \Omega_0 > \frac{4}{\sqrt{\pi}} \) the estimate above can be improved and we can prove that the profile \( | \gpm |^2 \) is exponentially small in \( \varepsilon \) inside the ``hole'', i.e. where \( \tfm \) is zero:

\begin{pro}[Exponential Smallness of the Density in the ``Hole'']
	\label{exp}
	\mbox{}	\\
    	Denote
    	\beq
        	\detf \equiv \left\{ \vec{r} \in \bo \: \big| \: r \leq \rtf - \varepsilon^{\frac{1}{3}} \right\}
    	\eeq
    	where \( \rtf \) is defined in \eqref{rtf}. For any \( \Omega_0 > \frac{4}{\sqrt{\pi}} \) and \( \varepsilon \) sufficiently small, there exist two constants \( C_{\Omega_0} \) and \( C_{\Omega_0}^{\prime} \) such that, for \( \vec{r} \in \detf \),     	
	\beq
    	\label{pointwise}
        	\big| \gpm(\vec{r}) \big|^2 \leq  C_{\Omega_0} \varepsilon^{\frac{1}{6}} \sqrt{|\log\varepsilon|} \: \exp\left[ -\frac{C^{\prime}_{\Omega_0} \mathrm{dist}(\vec{r}, \partial \detf)^2}{\varepsilon^{\frac{2}{3}}} \right].
    	\eeq
\end{pro}

The results stated above allow us to complete the proof of Proposition \ref{breaking}:

\begin{proof1}{Proposition \ref{breaking}}
	It remains to prove the statement for any $ \Omega(\varepsilon) = \Omega_0 / \varepsilon $. Suppose that the opposite statement is true, namely the ground state energy $ \gpe $ is reached on a symmetric vortex of the form $ \xi_n e^{in\vartheta} $. Then there must be some $ \bar{n}_{\varepsilon} \in \mathbb{N} $ such that
	\beq
	\label{symenebound1}
		E_{\bar{n}_{\varepsilon}}(\varepsilon) - \Omega(\varepsilon) \bar{n}_{\varepsilon} = \gpe \leq \frac{\tfe}{\varepsilon^2} + \frac{C |\log\varepsilon|}{\varepsilon}
	\eeq
	where we have used the upper bound for $ \gpe $ proved in Theorem \ref{energy}. Using the rough lower bound $ E_n(\varepsilon) \geq n^2 $, we immediately get the upper bound $ \bar{n}_{\varepsilon} \leq C/\varepsilon $ for some constant $ C $ independent of $ \varepsilon $. 
	\newline
	The right hand side of \eqref{symenebound1} can be bounded below by
	\bdm
		E_n - \Omega(\varepsilon)n \geq \int_{\bo} d\vec{r} \: \left[ \frac{n}{r} - \frac{\Omega(\varepsilon)r}{2} \right]^2 \xi_n^2(r) + \frac{\tff[\xi_n^2]}{\varepsilon^2} \geq  \int_{\bo} d\vec{r} \: \left[ n - \frac{\Omega(\varepsilon)r^2}{2} \right]^2 \xi_n^2(r) + \frac{\tfe}{\varepsilon^2}.
	\edm	
	Therefore one has the estimate
	\bdm
		\int_{\bo} d\vec{r} \: \left[ \bar{n}_{\varepsilon} - \frac{\Omega(\varepsilon)r^2}{2} \right]^2 \xi_{\bar{n}_{\varepsilon}}^2(r) \leq \frac{C |\log\varepsilon|}{\varepsilon}
	\edm
	Since $ \xi_{\bar{n}_{\varepsilon}} e^{i\bar{n}_{\varepsilon}\vartheta} $ is a ground state, it must satisfy the estimate \eqref{L1profile}, and then
	\bdm
		\frac{C |\log\varepsilon|}{\varepsilon} \geq \int_{\bo} d\vec{r} \: \left[ \bar{n}_{\varepsilon} - \frac{\Omega(\varepsilon)r^2}{2} \right]^2 \xi_{\bar{n}_{\varepsilon}}^2(r) \geq \int_{\bo} d\vec{r} \: \left[ \bar{n}_{\varepsilon} - \frac{\Omega(\varepsilon)r^2}{2} \right]^2 \tfm(r) - \frac{C \sqrt{|\log\varepsilon|}}{\varepsilon^{\frac{3}{2}}}
	\edm
	or
	\beq	
	\label{symenebound2}	
		\int_{\bo} d\vec{r} \: \left[ \bar{n}_{\varepsilon} - \frac{\Omega(\varepsilon)r^2}{2} \right]^2 \tfm(r) \leq  \frac{C_{\Omega_0} \sqrt{|\log\varepsilon|}}{\varepsilon^{\frac{3}{2}}}
	\eeq
	where we have used the bound
	\bdm
		\int_{\bo} d\vec{r} \: \left[ \bar{n}_{\varepsilon} - \frac{\Omega(\varepsilon)r^2}{2} \right]^2 \left( \xi_{\bar{n}_{\varepsilon}}^2(r) - \tfm(r) \right) \geq - \left\| \bar{n}_{\varepsilon} - \frac{\Omega(\varepsilon)r^2}{2} \right\|^2_{L^{\infty}(\bo)} \left\| \xi^2_{\bar{n}_{\varepsilon}} - \tfm \right\|_{L^1(\bo)} \geq - \frac{C \sqrt{|\log\varepsilon|}}{\varepsilon^{\frac{3}{2}}}.
	\edm
	The left hand side of \eqref{symenebound2} can be explicitly calculated: If $ \Omega_0 \leq 4/\sqrt{\pi} $, one has
	\bdm
		\int_{\bo} d\vec{r} \: \left[ \bar{n}_{\varepsilon} - \frac{\Omega(\varepsilon) r^2}{2} \right]^2 \tfm(r) = {\bar{n}_{\varepsilon}}^2 - \frac{\Omega(\varepsilon) \bar{n}_{\varepsilon}}{2} \left[ 1 + \frac{\pi \Omega_0^2}{48} \right] +  \frac{\Omega^2(\varepsilon)}{12} \left[ 1 +  \frac{\pi \Omega_0^2}{32} \right]
	\edm
	while, if $ \Omega_0 > 4/\sqrt{\pi} $,
	\bdm
		\int_{\bo} d\vec{r} \: \left[ \bar{n}_{\varepsilon} - \frac{\Omega(\varepsilon) r^2}{2} \right]^2 \tfm(r) =  {\bar{n}_{\varepsilon}}^2 - \Omega(\varepsilon) \bar{n}_{\varepsilon} \left[ 1 - \frac{4}{3\sqrt{\pi} \Omega_0} \right]  +  \frac{\Omega^2(\varepsilon) }{2} \left[ 1 - \frac{8}{3\sqrt{\pi}  \Omega_0} + \frac{8}{3\pi \Omega_0^2} \right].
	\edm
	By minimizing over $ \bar{n}_{\varepsilon} $, i.e., taking 
	\bdm
		\bar{n}_{\varepsilon} = \frac{\Omega(\varepsilon)}{4} \left[ 1 + \frac{\pi \Omega_0^2}{48} \right] 
	\edm
	in the first case and 
	\bdm
		\bar{n}_{\varepsilon} = \frac{\Omega(\varepsilon)}{2} \left[ 1 - \frac{4}{3\sqrt{\pi} \Omega_0} \right] 
	\edm
	in the second, we get
	\bdm	
		\int_{\bo} d\vec{r} \: \left[ \bar{n}_{\varepsilon} - \frac{\Omega(\varepsilon) r^2}{2} \right]^2 \tfm(r) \geq \frac{\Omega^2(\varepsilon)}{72} = \frac{\Omega_0^2}{72 \varepsilon^2} \equiv \frac{C^{\prime}_{\Omega_0}}{\varepsilon^2},
	\edm
	if $ \Omega_0 \leq 4/\sqrt{\pi} $, and 
	\bdm
		\int_{\bo} d\vec{r} \: \left[ \bar{n}_{\varepsilon} - \frac{\Omega(\varepsilon) r^2}{2} \right]^2 \tfm(r) \geq \frac{\Omega^2(\varepsilon)}{4} \left[ \frac{1}{3} + \frac{32}{9\pi \Omega_0^2} \right] \equiv \frac{C^{\prime}_{\Omega_0}}{\varepsilon^2},	
	\edm
	if $ \Omega_0 > 4/\sqrt{\pi} $. Therefore in both cases \eqref{symenebound2} implies that
	\bdm
		0 < C^{\prime}_{\Omega_0} \leq C_{\Omega_0} \sqrt{\varepsilon |\log\varepsilon|}
	\edm  
	for some strictly positive constant $ C^{\prime}_{\Omega_0} $. For $ \varepsilon $ sufficiently small this is a contradiction and then no symmetric vortex can be a ground state of the GP functional.
\end{proof1}

\subsection{The Regime $\Omega(\varepsilon)\gg 1/\varepsilon$}

In order to present the results in a transparent way we assume that the angular velocity is a power of \( 1/\varepsilon \), namely
\bdm
    	\Omega(\varepsilon) = \frac{\Omega_1}{\varepsilon^{1+\alpha}}
\edm
for some \( \alpha > 0 \)\footnote{Our analysis applies, in fact,  to arbitrary angular velocites \( \Omega(\varepsilon) \gg 1/\varepsilon \), one just has to replace $\Omega_1/\varepsilon^\alpha$ by $\varepsilon\Omega(\varepsilon)$.}. In this case the limiting functional is analogous to the one introduced in Section 2.3, provided  \( \Omega_0 \) is replaced by \( {\Omega_1}/{\varepsilon^{\alpha}} \) and the energy scale by \( \varepsilon^{2\alpha} \), i.e.
\beq
    	\label{tffa}
    	\tffa[\rho] \equiv \varepsilon^{2\alpha} \int_{\bo} d\vec{r} \: \left\{ \rho^2 - \frac{\Omega_1^2 r^2 \rho}{4 \varepsilon^{2\alpha}} \right\}.
\eeq
The ground state energy of this functional, i.e.,
\beq
\label{tfea}
    	\tfea \equiv \min_{    
		\bay
                	\rho \in \mathcal{D}^{TF} \\
                	\int \rho = 1
            	\eay} \tffa [\rho] = \tffa[\tfm_\varepsilon]
\eeq
is given by
\beq
	\label{tfea1}
    	\tfea = - \frac{\Omega_1^2}{4} \left( 1 - \frac{8 \varepsilon^{\alpha}}{3 \sqrt{\pi} \Omega_1} \right)
\eeq
and the corresponding minimizer is
\beq
    	\label{tfma}
    	\tfma = \frac{\Omega_1^2}{8 \varepsilon^{2\alpha}} \left[ r^2 - R_{\varepsilon}^2 \right]_+
\eeq
where
\beq
    	\label{rtfa}
    	\rtfa \equiv \sqrt{1 - \frac{4 \varepsilon^{\alpha}}{\sqrt{\pi} \Omega_1}}.
\end{equation}
Hence the function \( \tfma \) is supported in a very thin layer near the boundary and, as \( \varepsilon \rightarrow 0 \), it converges as a distribution to a radial delta function supported at \( r = 1 \): If 
 \( F(r) \) is a continuous function, one has
\bdm
	\int_{\bo} d\vec{r} \: \tfma(r) F(r) = \frac{\pi \Omega_1^2}{8 \varepsilon^{2\alpha}} \int_0^{1-\rtfa^2} dz \: z \: F\left(\sqrt{z+\rtfa^2}\right) = \int_0^1 dz \: z \: F\left(\sqrt{\frac{2\varepsilon^{\alpha}z}{\sqrt{\pi} \Omega_1 } + \rtfa^2}\right) \underset{\varepsilon \rightarrow 0}{\longrightarrow} F(1)
\edm
We can now state the main results for this regime, starting with the energy asymptotics for  \( \varepsilon \to 0 \):

\begin{teo}[Energy Asymptotics]
	\label{energya}
	\mbox{}	\\
    	For any \( \Omega_1 > 0 \), \( \alpha > 0 \), and for \( \varepsilon \) sufficiently small
    	\beq
	\label{energyaest}
        	\varepsilon^{2+2\alpha} \: \gpe = \tfea + \mbox{O}(\varepsilon^{2\alpha}) + \mbox{O}(\varepsilon^2 |\log \varepsilon|).
    	\eeq
\end{teo}

The two remainders in the asymptotic estimation of the GP energy \( \gpe \) have different sources: The second one,  of the order \( {|\log\varepsilon|}/{\varepsilon^{2\alpha}} \), is due to the convergence of the density to a delta function and the radial kinetic energy that is ignored in the TF functional. The other term, of order $1/\varepsilon^2$, is due to the approximation of the vortex structure in the region where the density is exponentially small by a trial function with a single vortex located at the origin. It is clear that the estimate  \eqref{energyest} is not the $\alpha\to 0$ limit of \eqref{energyaest}. We also note that for \( \alpha \geq 2 \)  the last error term in \eqref{energyaest} is larger than the second term in \eqref{tfea1}.

Since the function \( \tfma \) does not converge in any \( L^p \)-space, a result analogous to Corollary \ref{profile} does not hold. We are able to show, however, that the \( L^2 \)-norm of \( \gpm \) converges to zero almost everywhere, except for a thin region (with a size of order of a suitable power of \( \varepsilon \)) near the boundary. 

\begin{cor}[Density Asymptotics]
    	\label{profilea}
   	If  \( \Omega_1 > 0 \) and \( \varepsilon \) is sufficiently small,
    	\beq
        	\label{l2norma1}
        	\left\| \gpm \right\|^2_{L^2({\mathcal B}_{R_\varepsilon})} = O(\varepsilon^{\alpha}) + O(\varepsilon^{2-\alpha} |\log \varepsilon|)
    	\eeq
    	for \( 0 < \alpha < 2 \), while for  \( \alpha \geq 2 \),
    	\beq
        	\label{l2norma2}
        	\left\| \gpm \right\|^2_{L^2({\mathcal B}_{R_{\varepsilon,\beta}})} = O(\varepsilon^{2-\beta} |\log \varepsilon|)
    	\eeq
    	with $R_{\varepsilon, \beta}=(1-\varepsilon^\beta)^{1/2}$,  for any $1\leq\beta<2$.
\end{cor}

The above estimate is strengthened in the following proposition. The reason why we state Corollary 
\ref{profilea} separately is the analogy with the previous Corollary \ref{profile}. It is also used in the proof of the following.

\begin{pro}[Exponential Smallness of the Density]
	\label{expa}
	\mbox{}	\\
    	Denote
    	\beq
        	\detf' = \left\{ \vec{r} \in \bo \: \big| \: r \leq 1 - \varepsilon^{{\alpha^{\prime}}/{4}} \right\}
    	\eeq
    	and
    	\beq
        	\detf^{\prime\prime} = \left\{ \vec{r} \in \bo \: \big| \: r \leq 1 - \varepsilon^{{(2 - \beta)}/{4}} \right\}
    	\eeq
    	where \( \alpha^{\prime} = \min \left[ \alpha, 2-\alpha \right] \) and \( \beta \) is any number such that \( 1 \leq \beta < 2 \).
    	\newline
    	For any \( \Omega_1 > 0 \)  there exist constants \( C_{\Omega_1} \) and \( C_{\Omega_1}^{\prime} \), such that for  \( \varepsilon \) sufficiently small,
    	\beq
        	\big| \gpm(\vec{r}) \big|^2 \leq  C_{\Omega_1} \varepsilon^{{\alpha^{\prime}}/{3}} |\log \varepsilon | \: \exp\left[ -\frac{C^{\prime}_{\Omega_1} \mathrm{dist}(\vec{r}, \partial \detf')^2}{\varepsilon^{1+\frac{\alpha^{\prime}}{2}}} \right]
    	\eeq
    	if \( 0 < \alpha < 2 \) and \( \vec{r} \in \detf '\), and
    	\beq
        	\big| \gpm(\vec{r}) \big|^2 \leq  C_{\Omega_1} \varepsilon^{{(2-\beta)}/{3}} |\log \varepsilon| \: \exp\left[ -\frac{C^{\prime}_{\Omega_1} \mathrm{dist}(\vec{r}, \partial \detf^{\prime\prime})^2}{\varepsilon^{1+\frac{\alpha}{2}}} \right]
    	\eeq
    	if \( \alpha \geq 2 \) and \( \vec{r} \in \detf^{\prime\prime} \).
\end{pro}

A straightforward consequence of the above estimates together with the normalization of \( \gpm \) is that the density of any minimizer of the GP functional converges to \( \delta(1-r) \) in a distributional sense, in accord with the discussion in \cite{FiB}. 

Another important difference compared to the regime \( \Omega(\varepsilon) \sim 1/\varepsilon \) is the form of the trial function \eqref{triala} used in the proof of Theorem \ref{energya}.  This function is an eigenfunction of the angular momentum, i.e., the whole vorticity is concentrated at the origin. On the other hand, Proposition \ref{breaking} implies that the true minimizer cannot be an eigenfunction of the angular momentum, at least as long as \( \Omega(\varepsilon) \lesssim  1/\varepsilon \). Nevertheless we expect that the number of vortices contained in the region where \( \gpm \) is not exponentially small is negligible compared to the total vorticity of the function (see, e.g., the numerical simulations contained in \cite{KTU}). A wave function of this kind is often referred to in the physics literature as a ``giant vortex''. Since the vortex contribution to the energy depends essentially only on the winding number at the boundary of the thin region where the wave function of the condensate is not exponentially small,  a trial function with the vorticity concentrated at the origin can lead to a  good approximation to the energy.  
\newline
This behavior is also suggested by the fact that the minimization of a modified GP functional over the subspace of functions with fixed angular momentum with a subsequent minimization over the value of  the angular momentum gives a ground state energy with the same leading order asymptotics as \( \tfea \) and \( \gpe \). Indeed,  if we define (c.f. \cite{FiB})
\beq
	\label{tffgv}
    	{{\mathcal E}^{\rm TF}_{\varepsilon,\nu}}^{\prime}[\rho] \equiv \int_{\bo} d\vec{r} \: \left\{ \frac{\nu^2 \rho}{r^2} - \Omega_1 \nu \rho +  \varepsilon^{2\alpha} \rho^2 \right\}
\eeq
and
\beq
	\label{tfegv}
    	{\tfea}^{\prime} \equiv \min_{\nu \in \mathbb{R}^+} \min_{    
		\bay
                	\rho \in {\mathcal{D}^{TF}}^{\prime} \\
                	\int \rho = 1
            	\eay} {{\mathcal E}^{\rm TF}_{\varepsilon,\nu}}^{\prime} [\rho] 
\eeq	
where \( {\mathcal{D}^{TF}}^{\prime} \) is the natural domain for the functional \eqref{tffgv}, then it is not hard to see that
\beq
\label{giantvortex}
	\lim_{\varepsilon \to 0} {\tfea}^{\prime} = \lim_{\varepsilon \to 0} \tfea = \lim_{\varepsilon \to 0} \varepsilon^{2+2\alpha} \gpe = - \frac{\Omega_1^2}{4}
\eeq
even though \( {\tfea}^{\prime} > \tfea \) for any \( \varepsilon > 0 \). The functional \eqref{tffgv} is  obtained from \eqref{gpfmagnetic} by neglecting the radial part of the kinetic energy and restricting it to eigenfunctions with fixed angular momentum\footnote{More precisely, the correct value of the angular momentum should be the integer part of \( \nu / \varepsilon^{1+\alpha} \) but the difference between these two quantities produces a correction of smaller order in \eqref{tfegv}.} \( \nu/\varepsilon^{1+\alpha} \). In other words the TF functional \eqref{tffgv} describes the asymptotic behavior of the GP functional almost as well as \eqref{tffa}. 

However,  while the heuristic discussion in \cite{FiB} suggests that  a giant vortex occurs only for  angular velocities larger than \( 1/(\varepsilon^2 |\log\varepsilon|) \), we found no evidence in our rigorous analysis that a change in the minimizer occurs above that threshold. In fact such a critical angular velocity for the transition to the giant vortex is estimated in \cite{FiB} by imposing the condition that the ground state energy \( {\tfea}^{\prime} \) equals, to leading order, a certain upper bound for \( \varepsilon^{2+2\alpha} \gpe \). This upper bound, however, is calculated in \cite{FiB} with a trial function of the form \eqref{trial} and  is not optimal. In fact,  the same comparison with a better upper bound\footnote{For instance the one calculated using the trial function \eqref{triala} in the proof of Theorem \ref{energya}.} for \( \varepsilon^{2+2\alpha} \gpe \) gives the correct answer, namely that \( \varepsilon^{2+2\alpha} \gpe \) is close to \( {\tfea}^{\prime} \) for any \( \alpha > 0 \) (see \eqref{giantvortex}). Hence the transition to the giant vortex should occur for any angular velocity \( \Omega(\varepsilon) \) of order higher than \( 1/\varepsilon \).

\section{Proofs}

\subsection{The Regime $\Omega(\varepsilon)\sim1/\varepsilon$}

The main result concerning the regime $\Omega(\varepsilon)\sim1/\varepsilon$ is Theorem \ref{energy} and we start by proving it. Some technical but crucial details of the proof are contained in Subsection \ref{3.1.1} (Proposition \ref{normpro} and Theorem \ref{remainder}), where we present some estimates for the kinetic energy of the trial function.

\begin{proof1}{Theorem \ref{energy}}
	We are going to prove the result by comparing an upper bound for the ground state energy with a suitable lower bound.

	\emph{Lower Bound:}
   	The lower bound for \( \gpe \) is actually trivial. By simply neglecting the positive contribution of the magnetic kinetic energy in \eqref{gpfmagnetic} we immediately get
    	\beq
    	\label{lbound}
        	\gpf[\Psi] \geq \frac{\tff[|\Psi|^2]}{\varepsilon^2} \geq \frac{\tfe}{\varepsilon^2}.
    	\eeq

   	\emph{Upper Bound:}
    	We prove the upper bound by testing the functional on a trial function of the following form
    	\beq
    	\label{trial}
        	\trial(\vec{r}) = c_{\varepsilon} f_{\varepsilon}(r) \cut(\vec{r}) g_{\varepsilon}(\vec{r}).
    	\eeq
    	The radial part is given by
    	\beq
        	f_{\varepsilon}(r) =
        	\left\{
        	\begin{array}{ll}
            		\sqrt{\tfm}     &   \mbox{if} \:\:\:\: \Omega_0 \leq \displaystyle{\frac{4}{\sqrt{\pi}}}    \\
            		\mbox{}     &   \mbox{} \\
            		j_{\varepsilon} \sqrt{\tfm}     &   \mbox{if} \:\:\:\: \Omega_0 > \displaystyle{\frac{4}{\sqrt{\pi}}}
        	\end{array}
        	\right.
    	\eeq
    	where \( j_{\varepsilon} \) is a suitable cut-off function to regularize \( \sqrt{\tfm} \) at the boundary of the hole. Our choice is
	    	\beq
    	\label{cutprof}
        	j_{\varepsilon}(r) = \left\{
        	\begin{array}{ll}
            		0   &   \mbox{if} \:\:\:\: r \leq \rtf  \\
            		\mbox{} &   \mbox{} \\
            		\displaystyle{\frac{r-\rtf}{\varepsilon}}   &   \mbox{if} \:\:\:\: \rtf \leq r \leq \rtf + \varepsilon  \\
            		\mbox{} &   \mbox{} \\
            		1   &   \mbox{otherwise}.
        	\end{array}
        	\right.
    	\eeq
    	The function \( g_{\varepsilon} \) is a phase factor that can be expressed in complex coordinates \( z = x+iy \) as
    	\beq
    	\label{phase}
        	g_{\varepsilon}(z) = \prod_{i \in \latt} \frac{z - z_i}{|z- z_i|}
    	\eeq
    	where \( \latt \) is a square lattice of spacing \( \spac \) defined in the following way:
    	\beq
    	\label{lattice}
        	\latt = \left\{ \vec{r} = (m \spac, n \spac), \: \: m,n \in \mathbb{Z} \: \bigg| \: r \leq 1 - 2 \sqrt{2} \spac \right\}.
    	\eeq
    	We assume that the spacing is of order \( \sqrt{\varepsilon} \), i.e. \( \spac = \delta \sqrt{\varepsilon} \), for some \( \delta > 0 \) independent of \( \varepsilon \), so that the number of lattice points, denoted by \( N_{\varepsilon} \), is proportional to \( 1/\varepsilon \).
    	\newline
    	We note that the phase \( g_{\varepsilon} \) carries vortices of degree 1 centered at the lattice points. Moreover, each vortex core is contained inside the fundamental cell and its radius is of order \( \sqrt{\varepsilon} \). This choice is suggested by previous works (see e.g. \cite{Beth2,Igna1,Igna2}) on rotating condensates, where it is shown that vortices of higher degree than 1 are energetically unfavorable.
    	\newline
    	Since \( g_{\varepsilon} \) is not differentiable at the points of the lattice  we need to multiply it by a function, \( \cut \),  that vanishes at these points and we take
    	\beq
        	\cut(\vec{r}) =
        	\left\{
        	\begin{array}{ll}
            		1   &   \mbox{if} \:\:\:\: |\vec{r} - \vec{r}_i| \geq \varepsilon^{\eta}  \\
            		\mbox{} &   \mbox{} \\
            		\displaystyle{\frac{|\vec{r} - \vec{r}_i|}{\varepsilon^{\eta}}}   &   \mbox{if} \:\:\:\: |\vec{r} - \vec{r}_i| \leq \varepsilon^{\eta}
        		\end{array}
        	\right.
    	\eeq
    	for some \( \eta > 1/2 \).
    	\newline
    	Finally the constant \( c_{\varepsilon} \) is fixed by the normalization condition and it can be easily checked that, for \( \varepsilon \) sufficiently small,
    	\beq
    	\label{constant}
        	1 \leq c^2_{\varepsilon} \leq 1+ C \varepsilon.
    	\eeq
    	Setting
    	\beq
    	\label{region}
        	\Lambda = \bo \backslash \bigcup_{i \in \latt} \bi
    	\eeq
    	where \( \bi \) is a ball of radius \( \varepsilon^{\eta} \) centered at \( \vec{r}_i \), the functional evaluated on the trial function \eqref{trial} is given by
    	\bdm
        	\gpf[\trial] = c_{\varepsilon}^2 \int_{\Lambda} d\vec{r} \: \left| \nabla f_{\varepsilon} \right|^2 + c_{\varepsilon}^2 \int_{\Lambda} d\vec{r} \: f_{\varepsilon}^2 \: \left| \left( \nabla - i \vec{A_{\varepsilon}} \right) g_{\varepsilon} \right|^2 +
    	\edm
    	\bdm
        	+ \int_{\cup_{i \in \latt} \bi} d\vec{r} \: \left| \left( \nabla - i \vec{A_{\varepsilon}} \right) \trial \right|^2 + \frac{\tff [ | \trial |^2 ]}{\varepsilon^2} =
    	\edm
    	\bdm
        	= c_{\varepsilon}^2 \int_{\Lambda} d\vec{r} \: \left| \nabla f_{\varepsilon} \right|^2 + c_{\varepsilon}^2 \int_{\Lambda} d\vec{r} \: f_{\varepsilon}^2 \: \left| \left( \nabla - i \vec{A_{\varepsilon}} \right) g_{\varepsilon} \right|^2 + c_{\varepsilon}^2 \int_{\cup_{i \in \latt} \bi} d\vec{r} \: \left| \nabla \left( \chi_{\varepsilon} f_{\varepsilon} \right) \right|^2 +
    	\edm
    	\bdm
        	+  c_{\varepsilon}^2 \int_{\cup_{i \in \latt} \bi} d\vec{r} \:  \cut^2 f_{\varepsilon}^2 \left| \left( \nabla - i \vec{A_{\varepsilon}} \right) g_{\varepsilon} \right|^2 + \frac{\tff [ | \trial |^2 ]}{\varepsilon^2} \leq
    	\edm
    	\bdm
        	\leq  c_{\varepsilon}^2 \int_{\bo} d\vec{r} \: \left| \nabla f_{\varepsilon} \right|^2 + c_{\varepsilon}^2 \int_{\Lambda} d\vec{r} \: f_{\varepsilon}^2 \: \left| \left( \nabla - i \vec{A_{\varepsilon}} \right) g_{\varepsilon} \right|^2 +  c_{\varepsilon}^2 \int_{\cup_{i \in \latt} \bi} d\vec{r} \: \cut^2 f_{\varepsilon}^2 \left| \left( \nabla - i \vec{A_{\varepsilon}} \right) g_{\varepsilon} \right|^2 +
    	\edm
    	\bdm
        	+ \frac{\tff [ | \trial |^2 ]}{\varepsilon^2} + \frac{C}{\varepsilon} \leq
    	\edm
    	\bdm
    	    \leq  c_{\varepsilon}^2 \int_{\bo} d\vec{r} \: \left| \nabla f_{\varepsilon} \right|^2 + C_1 \int_{\Lambda} d\vec{r} \: \left| \left( \nabla - i \vec{A_{\varepsilon}} \right) g_{\varepsilon} \right|^2 + C_2 \int_{\cup_{i \in \latt} \bi} d\vec{r} \: \cut^2 \left| \nabla g_{\varepsilon} \right|^2 +
    	\edm
    	\bdm
        	+ C_3 \int_{\cup_{i \in \latt} \bi} d\vec{r} \: \left| \magnp \right|^2 + \frac{\tff [ | \trial |^2 ]}{\varepsilon^2} + \frac{C}{\varepsilon} \leq
    	\edm
    	\bdm
        	\leq  c_{\varepsilon}^2 \int_{\bo} d\vec{r} \: \left| \nabla f_{\varepsilon} \right|^2 +  C_{1} \int_{\Lambda} d\vec{r} \: \left| \left( \nabla - i \vec{A_{\varepsilon}} \right) g_{\varepsilon} \right|^2 +  C_2 \int_{\cup_{i \in \latt} \bi} d\vec{r} \: \cut^2 \left| \nabla g_{\varepsilon} \right|^2 +
    	\edm
    	\bdm
        	+ \frac{\tff [ | \trial |^2 ]}{\varepsilon^2} + \frac{C_4}{\varepsilon^{3-4\eta}} + \frac{C}{\varepsilon}
    	\edm
    	where we have used the uniform boundedness of \( f_{\varepsilon} \), the estimate \eqref{constant}, and the fact that the number of lattice points \( N_{\varepsilon} \) is bounded by \( C/\varepsilon \).
    	\newline
    	The gradient of the phase \( g_{\varepsilon} \) can be bounded from above inside any ball \( \bi \):
    	\beq
    	\label{gradient}
        	\left| \nabla g_{\varepsilon} \right| \leq \sum_{j \in \latt} \frac{1}{|\vec{r} - \vec{r_j}|} \leq \frac{1}{|\vec{r} - \vec{r_i}|} + \frac{N_{\varepsilon}}{\inf_{j \neq i} |\vec{r} - \vec{r}_j|} \leq \frac{1}{|\vec{r} - \vec{r_i}|} + \frac{N_{\varepsilon}}{\spac}
    	\eeq
    	for any \( \vec{r} \in \bi \), so that
    	\bdm
        	\int_{\cup_{i \in \latt} \bi} d\vec{r} \: \cut^2 \left| \nabla g_{\varepsilon} \right|^2 \leq \frac{ C \left| \cup_{i \in \latt} \bi \right|}{\varepsilon^{2\eta}} + \frac{C^{\prime} \left| \cup_{i \in \latt} \bi \right|}{\varepsilon^3} \leq \frac{C}{\varepsilon} + \frac{C^{\prime}}{\varepsilon^{4-2\eta}}
    	\edm
    	and hence
    	\bdm
        	\gpf[\trial] \leq \int_{\bo} d\vec{r} \: \left| \nabla f_{\varepsilon} \right|^2 +  C_{1} \int_{\Lambda} d\vec{r} \: \left| \left( \nabla - i \vec{A_{\varepsilon}} \right) g_{\varepsilon} \right|^2  + \frac{\tff [ | \trial |^2 ]}{\varepsilon^2} + \frac{C_2}{\varepsilon}
    	\edm
    	for any \( \eta > 3/2 \).
    	\newline
    	Moreover, the radial part of the kinetic energy can be bounded by a constant  if \( \Omega_0 \leq \frac{4}{\sqrt{\pi}} \), and by
    	\bdm
        	\int_{\bo} d\vec{r} \: \left| \nabla f_{\varepsilon} \right|^2 \leq C_1 \int_{\bo} d\vec{r} \: \left| \nabla j_{\varepsilon} \right|^2 + C_2 \int_{R_0}^{1} dr \: r \: \frac{j_{\varepsilon}^2}{\tfm} \leq C_3 + C_4 \int_{R_0+\varepsilon}^{1} dr \: \frac{r}{r^2- R_0^2} \leq C | \log \varepsilon|
    	\edm
    	if \( \Omega_0 > \frac{4}{\sqrt{\pi}} \). Then we get
    	\bdm
        	\gpf[\trial] \leq C_{1} \int_{\Lambda} d\vec{r} \: \left| \left( \nabla - i \vec{A_{\varepsilon}} \right) g_{\varepsilon} \right|^2  + \frac{\tff [ | \trial |^2 ]}{\varepsilon^2} + \frac{C_2}{\varepsilon}
    	\edm
    	for a possibly different constant \( C_2 \).
    	\newline
    	The upper bound\footnote{Note that the constant \( C_{\Omega_0} \) actually depends linearly on \( \eta \) (see the proof of Lemma \ref{kinetic}, in particular Eq.\ \eqref{innerbd}).} now follows using Proposition \ref{normpro} and Theorem \ref{remainder} in the next section, choosing \( \delta = \sqrt{\frac{2\pi}{\Omega_0}} \) and \( 5/2 < \eta < \infty \):
    	\beq
    	\label{ubound}
        	\gpf[\trial] \leq \frac{\tfe}{\varepsilon^2} + \frac{C_{\Omega_0} |\log \varepsilon|}{\varepsilon}.
    	\eeq
\end{proof1}

\begin{rem}[Dirichlet Problem]
	\label{Dirichlet}
	\mbox{}	\\
	\emph{The proof in the case of Dirichlet boundary conditions, i.e.,  if \eqref{domain} is replaced with \( H^1_0(\bo) \), looks exactly the same: The trial function has simply to be multiplied by a cut-off function, which is \( 1 \) everywhere except for a very thin region in the neighborhood of the boundary where it goes to \( 0 \), in order to satisfy the required boundary conditions. The error coming from such a cut-off function can then be included in the remainder in \eqref{ubound}.}
\end{rem}

\begin{proof1}{Corollary \ref{profile}}
	Let us first consider the case \( \Omega_0 < \frac{4}{\sqrt{\pi}} \). Using the explicit form of the TF minimizer \( \tfm \) (see \eqref{tfm}) and the estimate \eqref{ubound}, one can calculate
    	\bdm
        	\int_{\bo} d\vec{r} \: \left( |\gpm|^2 - \tfm \right)^2 = \tff[|\gpm|^2] - \frac{2}{\pi} + \frac{\Omega_0^2}{8} + \int_{\bo} d\vec{r} \: \left( \tfm \right)^2 \leq
    	\edm
    	\bdm
        	\leq \tfe - \frac{2}{\pi} + \frac{\Omega_0^2}{8} + \int_{\bo} d\vec{r} \: \left( \tfm \right)^2 + C_{\Omega_0} \varepsilon |\log \varepsilon| = C_{\Omega_0} \varepsilon |\log \varepsilon|
    	\edm
    	and then the \(L^1-\)bound follows from Schwarz's inequality.
    	\newline
    	On the other hand, if \( \Omega_0 \geq \frac{4}{\sqrt{\pi}} \), one has
    	\bdm
        	\int_{\dtf} d\vec{r} \: \left( |\gpm|^2 - \tfm \right)^2 = \tff[|\gpm|^2] - \tfe - \int_{\bo \setminus \dtf} d\vec{r} \: | \gpm |^4 +
    	\edm
    	\bdm
        	+ \frac{\Omega_0^2}{4} \int_{\bo \setminus \dtf} d\vec{r} \: (r^2 - \rtf^2) |\gpm|^2 \leq  \tff[|\gpm|^2] - \tfe \leq  C_{\Omega_0} \varepsilon |\log \varepsilon|
    	\edm
    	where we have used again \eqref{tfm} and \eqref{ubound}. From the same inequality one also has
    	\bdm
		\int_{\dtf} d\vec{r} \: \left( |\gpm|^2 - \tfm \right)^2 + \int_{\bo \setminus \dtf} d\vec{r} \: | \gpm |^4 = \tff[|\gpm|^2] - \tfe + \frac{\Omega_0^2}{4} \int_{\bo \setminus \dtf} d\vec{r} \: (r^2 - \rtf^2) |\gpm|^2 \leq
	\edm
	\bdm
		\leq \tff[|\gpm|^2] - \tfe \leq  C_{\Omega_0} \varepsilon |\log \varepsilon|,
	\edm
	so that
	\beq
    	\label{l4norm}
        	\int_{\bo \setminus \dtf} d\vec{r} \: | \gpm |^4 \leq C_{\Omega_0} \varepsilon |\log \varepsilon|.
    	\eeq
\end{proof1}

\begin{proof1}{Proposition \ref{exp}}
	The proof is similar to the one of Proposition 2.5 in \cite{Afta1}.
    	\newline
    	The variational equation satisfied by \( \gpm \) is
    	\beq
    	\label{variational}
        	- \Delta \gpm - \frac{\Omega_0}{\varepsilon} L \gpm + \frac{2}{\varepsilon^2} |\gpm|^2 \gpm = \chem \gpm
    	\eeq
    	where the chemical potential \( \chem \) is fixed by the \( L^2-\)normalization of \( \gpm \):
    	\beq
    	\label{chem}
        	\chem = \gpe + \frac{\| \gpm \|_4^4}{\varepsilon^2}.
    	\eeq
    	Setting \( U_{\varepsilon} \equiv | \gpm |^2 \) and using the simple estimate
    	\bdm
        	\frac{\Omega_0}{\varepsilon} \left| {\gpm}^* L \gpm \right| \leq \left| \nabla \gpm \right|^2 + \frac{\Omega_0^2 r^2 |\gpm|^2}{4 \varepsilon^2}
    	\edm
    	one can easily check that
    	\bdm
        	-\frac{1}{2} \Delta U_{\varepsilon} \leq \left[ \frac{\Omega_0^2 r^2}{4} + \varepsilon^2 \chem - 2 U_{\varepsilon} \right] \frac{U_{\varepsilon}}{\varepsilon^2}.
    	\edm
    	Moreover, thanks to Theorem \ref{energy} and Corollary \ref{profile},
    	\bdm
        	\varepsilon^2 \chem \leq \tfe + \| \tfm \|_2^2 + C_{\Omega_0} \sqrt{\varepsilon | \log \varepsilon |}
    	\edm
    	so that
    	\bdm
        	-\frac{1}{2} \Delta U_{\varepsilon} \leq \left[ \frac{\Omega_0^2 (r^2 - \rtf^2)}{4} - 2 U_{\varepsilon} + C_{\Omega_0} \sqrt{\varepsilon | \log \varepsilon |}  \right] \frac{U_{\varepsilon}}{\varepsilon^2}.
    	\edm
    	If we define
    	\bdm
        	\tilde \detf \equiv \left\{ \vec{r} \in \bo \: \Big| \: r \leq \rtf - \frac{\varepsilon^{\frac{1}{3}}}{2} \right\}
    	\edm
    	then, for \( \varepsilon \) sufficiently small, the function \( U_{\varepsilon} \) is subharmonic in \( {\tilde \detf} \) and therefore, for any point \( \vec{r} \in {\tilde \detf} \) with  \( \mathrm{dist}(\vec{r}, \partial {\tilde \detf}) \geq \varrho \),
    	\bdm
        	U_{\varepsilon}(\vec{r}) \leq \frac{1}{\pi \varrho^2} \int_{\mathcal{B}_{\varrho}(\vec{r})} d\vec{x} \: U_{\varepsilon} (\vec{x}) \leq \frac{C \| U_{\varepsilon} \|_{L^2(\bo \setminus \dtf)}}{\varrho}.
    	\edm
    	Hence, using the estimate \eqref{l4norm} and choosing, for instance, \( \varrho = \varepsilon^{\frac{1}{3}}/2 \), we can conclude that
    	\bdm
        	U_{\varepsilon}(\vec{r}) \leq C_{\Omega_0} \varepsilon^{\frac{1}{6}} \sqrt{| \log \varepsilon |}
    	\edm
    	for any \( \vec{r} \in \detf \).
    	\newline
    	Let us now define
    	\bdm
        	U_{\varepsilon}^{\prime} \equiv \frac{U_{\varepsilon}}{C_{\Omega_0} \varepsilon^{\frac{1}{6}} \sqrt{| \log \varepsilon |}}.
    	\edm
    	For any \( \vec{r} \in \detf \),
    	\bdm
        	- \Delta U^{\prime}_{\varepsilon} + \frac{C U^{\prime}_{\varepsilon}}{\varepsilon^{\frac{4}{3}}} \leq 0
    	\edm
    	and
    	\bdm
        	0 \leq U_{\varepsilon}^{\prime} \leq 1
    	\edm
    	i.e.,  \( U_{\varepsilon}^{\prime} \) is a subsolution in \( \detf \) of
    	\bdm
        	\left\{
        	\begin{array}{l}
            		- \Delta u + \displaystyle{\frac{C u}{\varepsilon^{\frac{4}{3}}}} = 0   \\
            		\mbox{} \\
            		u(\partial \detf) = 1.
        	\end{array}
        	\right.
    	\edm
    	On the other hand it is not so hard to verify (see e.g. Lemma 2 in \cite{Beth1}) that the function
    	\bdm
        	\exp \left\{ \frac{\sqrt{C} \left[ r^2- ( \rtf - \varepsilon^{\frac{1}{3}} )^2 \right]}{4 \varepsilon^{\frac{2}{3}} ( \rtf - \varepsilon^{\frac{1}{3}} )}  \right\}
    	\edm
    	is a supersolution for the same problem if
    	\bdm
        	\varepsilon^{\frac{2}{3}} \leq \frac{3 ( \rtf - \varepsilon^{\frac{1}{3}} )}{4}
    	\edm
    	and hence for any \( \varepsilon \) sufficiently small. The result now follows from the comparison principle.
\end{proof1}

\subsubsection{Technical Estimates}\label{3.1.1}

In this section we want to present some estimates involving the function \eqref{trial}. We start by stating a simple but important result:
	
\begin{pro}[Upper Bound on the TF Energy]
    	\label{normpro}
	\mbox{}	\\
    	Let \( \trial \) be the function defined in \eqref{trial}, with \( \eta > 1 \) and \( \Omega_0 > 0 \), then, for \( \varepsilon \) sufficiently small,
    	\beq
    	\label{thomas}
        	\tff[|\trial|^2] \leq \tfe + C_{\Omega_0} \varepsilon.
    	\eeq
\end{pro}

\begin{proof}
    	A simple estimate using \eqref{constant} shows that
    	\bdm
        	\tff[|\trial|^2]- \tff[\tfm]  \leq \int_{\Lambda} d\vec{r} \: \left\{ \left[ c_{\varepsilon}^4 f_{\varepsilon}^4 - (\tfm)^2 \right] - \frac{\Omega_0^2 r^2 \left[ c_{\varepsilon}^2 f_{\varepsilon}^2 - \tfm \right]}{4} \right\} +
    	\edm
    	\bdm
        	+ C \varepsilon^{2\eta - 1} \leq C^{\prime} \left\| c_{\varepsilon}^2 f_{\varepsilon}^2 - \tfm \right\|_{L^1(\bo)} + C \varepsilon^{2\eta - 1}
    	\edm
    	where \( \Lambda \) is the region defined in \eqref{region} and we have used the fact that \( \tfm \) and \( \trial \) are both uniformly bounded and the area \( \left| \cup_{i \in \latt} \bi \right|$ is bounded by $C \varepsilon^{2\eta - 1} \).
    	\newline
    	Now
    	\bdm
        	\left\| c_{\varepsilon}^2 f_{\varepsilon}^2 - \tfm \right\|_{L^1(\bo)} \leq \int_{\Lambda} d\vec{r} \: \left| \tfm - c^2_{\varepsilon} f^2_{\varepsilon} \right| + C \varepsilon^{2\eta -1}
    	\edm
    	and, setting \( d_{\varepsilon} = c^2_{\varepsilon} - 1 \leq C \varepsilon \), the first term on the right-hand side is bounded by
    	\bdm
        	d_{\varepsilon} \int_{\bo} d\vec{r} \: \tfm \leq C_{\Omega_0} \varepsilon
    	\edm
    	if \( \Omega_0 \leq \frac{4}{\sqrt{\pi}} \). On the other hand, if \( \Omega_0 \geq \frac{4}{\sqrt{\pi}} \),
    	\bdm
        	\int_{\Lambda} d\vec{r} \: \left| \tfm - c^2_{\varepsilon} f^2_{\varepsilon} \right| \leq d_{\varepsilon} \int_{\dtf} d\vec{r} \: \tfm + C_{\Omega_0} \varepsilon
    	\edm
    	where the last term is due to the cut-off function \( j_{\varepsilon} \) and \( \dtf \) is defined in \eqref{dtf}.
\end{proof}

The main result contained in this Section is the following

\begin{teo}[Upper Bound on the Vortex Contribution]
	\label{remainder}
	\mbox{}	\\
    	Let \( g_{\varepsilon} \) be the function defined in \eqref{phase}, \( \spac = \delta \sqrt{\varepsilon} \) and \( \Omega_0 > 0 \). There exists a constant \( C_{\Omega_0, \delta} \) independent of \( \varepsilon \) such that for \( \varepsilon \) sufficiently small and \( 5/2 < \eta < \infty \)
    	\beq
    	\label{phasees}
        	\int_{\Lambda} d\vec{r} \: \left| \left( \nabla - i \vec{A_{\varepsilon}} \right) \: g_{\varepsilon} \right|^2 \leq \frac{\pi}{2\varepsilon^2}\left(\frac{\Omega_0}{2} - \frac{\pi}{\delta^2} \right)^2 + \frac{C_{\Omega_0,\delta} |\log\varepsilon|}{\varepsilon}
    	\eeq
    	where \( \Lambda \) (depending on $\varepsilon^\eta $) is defined in \eqref{region}.
\end{teo}

\begin{rem}[Vortex Lattice]
	\mbox{}	\\
	\emph{As far as the leading order of the GP energy is concerned, the vortex structure of the minimizer \( \gpm \) is not so important: The choice of a regular square lattice in \eqref{lattice} is just the simplest for computational purposes but the result in Theorem \ref{remainder} is expected to hold for any trial function with vortices on a regular lattice, provided that \( \delta^2 \) in \eqref{phasees} is replaced with the volume of the rescaled fundamental cell, which is the relevant parameter in the estimate.}
\end{rem}

\begin{proof}
	Expanding the expression in \eqref{phasees}, we get
    	\beq
    	\label{decomp}
        	\int_{\Lambda} d\vec{r} \: \left| \left( \nabla - i \vec{A_{\varepsilon}} \right) g_{\varepsilon} \right|^2 =  \int_{\Lambda} d\vec{r} \: \left| \nabla g_{\varepsilon} \right|^2 + \frac{i\Omega_0}{\varepsilon} \int_{\Lambda} d\vec{r} \: g_{\varepsilon}^* \left( \vec{r} \times \nabla g_{\varepsilon} \right) +
    	\eeq
    	\bdm
        	+ \frac{\Omega_0^2}{4 \varepsilon^2} \int_{\Lambda} d\vec{r} \: r^2 |g_{\varepsilon}|^2.
    	\edm
    	The last term can be easily bounded from above by
    	\bdm
        	\frac{\Omega_0^2}{4 \varepsilon^2} \int_{\bo} d\vec{r} \: r^2 = \frac{\pi \Omega_0^2}{8 \varepsilon^2}.
    	\edm
    	Using the fact that \( g_{\varepsilon} = e^{i \phi} \), where
    	\beq
    	\label{phase1}
        	\phi(\vec{r}) = \sum_{i \in \latt} \arctan \left[ \frac{y-y_i}{x-x_i} \right]
    	\eeq
    	the second term can be explicitly calculated: By applying Stokes's theorem,
    	\bdm
        	\frac{i\Omega_0}{\varepsilon} \int_{\Lambda} d\vec{r} \: g_{\varepsilon}^* \left( \vec{r} \times \nabla g_{\varepsilon} \right) = - \frac{\Omega_0}{\varepsilon} \int_{\Lambda} d\vec{r} \: \vec{r} \times \nabla \phi = - \frac{\Omega_0}{2\varepsilon} \int_{\Lambda} d\vec{r} \: \nabla \times \left( r^2 \nabla \phi \right) =
    	\edm
    	\bdm
        	= - \frac{\Omega_0}{2\varepsilon} \int_{\partial \bo} d \vec{s} \: \cdot \: \nabla \phi + \frac{\Omega_0}{2\varepsilon} \sum_{i \in \latt} \int_{\partial \bi} d\vec{s} \: \cdot \: r^2 \nabla \phi =
    	\edm
    	\beq
    	\label{rotation1}
        	= - \frac{\pi \Omega_0 N_{\varepsilon}}{\varepsilon} + \frac{\pi \Omega_0}{\varepsilon} \sum_{i\in \latt} r_i^2 + \frac{\Omega_0}{2\varepsilon} \sum_{i \in \latt} \int_{\partial \bi} d\vec{s} \: \left( r^2-r_i^2 \right) \cdot \: \nabla \phi.
    	\eeq
    	Since for any \( \vec{r} \in \partial \bi \),
    	\bdm
        	\left| r^2 - r_i^2 \right| \leq C \varepsilon^{\eta}
    	\edm
    	the last term in the expression above can easily be bounded by
    	\bdm
        	\left| \frac{\Omega_0}{2\varepsilon} \sum_{i \in \latt} \int_{\partial \bi} d\vec{s} \: \left( r^2-r_i^2 \right) \cdot \: \nabla \phi \right| \leq \frac{C_{\Omega_0} N_{\varepsilon}}{\varepsilon^{1-\eta}} \int_{\partial \bi} d\vec{s} \: \left|  \nabla \phi \right| \leq \frac{C_{\Omega_0} N^2_{\varepsilon}}{\varepsilon^{1-\eta}}
    	\edm
    	where we have used the estimate \eqref{gradient}
    	\beq
    	\label{nabla}
        	\left| \nabla \phi (\vec{r}) \right| \leq \sum_{i \in \latt} \frac{1}{|\vec{r} - \vec{r_i}|} \leq \frac{N_{\varepsilon}}{\inf_{i \in \latt} |\vec{r} - \vec{r}_i|} \leq \frac{N_{\varepsilon}}{\varepsilon^{\eta}}.
    	\eeq
    	Since the lattice spacing \( \spac \) is chosen to be equal to \( \delta \sqrt{\varepsilon} \), the number of lattice points satisfies the bound
    	\beq
	\label{numberu}
        	N_{\varepsilon} \leq \frac{\pi \left( 1-\frac{3\spac}{\sqrt{2}} \right)^2}{\spac^2} \leq \frac{C_{\delta}}{\varepsilon}
    	\eeq
    	and then
    	\beq
    	\label{rot3}
        	\left| \frac{\Omega_0}{2\varepsilon} \sum_{i \in \latt} \int_{\partial \bi} d\vec{s} \: \left( r^2-r_i^2 \right) \cdot \: \nabla \phi \right| \leq \frac{C_{\Omega_0,\delta}}{\varepsilon^{3-\eta}}.
    	\eeq
    	Moreover the sum appearing in \eqref{rotation1} can be replaced by the integral over \( \bo \): let \( \cell \) and \( \celli \) be the fundamental cell centered at the origin and at \( \vec{r}_i \) respectively,
    	\bdm
        	r_i^2 - \frac{1}{\spac^2} \int_{\celli} d\vec{r} \: r^2 = \frac{1}{\spac^2} \int_{\cell} d\vec{r} \: r^2 = \frac{\spac^2}{6}
    	\edm
    	so that, setting \( \mathcal{A}_{\varepsilon} \equiv \bo \setminus ( \cup_{i \in \latt} \celli) \),
    	\bdm
        	\sum_{i \in \latt} \: r_i^2 = \frac{1}{\spac^2} \int_{\cup_{i \in \latt} \celli} d\vec{r} \: r^2 + \frac{N_{\varepsilon} \spac^2}{6} \leq \frac{1}{\spac^2} \int_{\bo} d\vec{r} \: r^2 - \frac{1}{\spac^2} \int_{\mathcal{A}_{\varepsilon}} d\vec{r} \: r^2  + C \leq
	\edm
	\beq
	\label{rot2}
		\leq \frac{\pi}{2 \spac^2} - \frac{(1-C^{\prime}\spac)^2 (\pi - N_{\varepsilon} \spac^2)}{\spac^2} + C \leq - \frac{\pi}{2 \spac^2} + N_{\varepsilon} + \frac{C^{\prime}(\pi - N_{\varepsilon} \spac^2)}{\spac} + C
	\eeq
	because the lattice is chosen in such a way that, for any \( i \in \latt \), \( r_i \leq 1 - 2\sqrt{2} \spac \).
    	\newline
    	From inequalities \eqref{rot3} and \eqref{rot2} we then get (for any \( \eta > 5/2 \))
    	\bdm
    	    \frac{i\Omega_0}{\varepsilon} \int_{\Lambda} d\vec{r} \: g_{\varepsilon}^* \left( \vec{r} \times \nabla g_{\varepsilon} \right) \leq - \frac{\pi^2 \Omega_0}{2 \varepsilon \spac^2} + \frac{C^{\prime}_{\Omega_0,\delta}(\pi - N_{\varepsilon} \spac^2)}{\varepsilon \spac} + \frac{C_{\Omega_0}}{\varepsilon}
    	\edm
    	but the number of points in the lattice can be estimated below  as     	\beq
	\label{number}
    	    \left| N_{\varepsilon} - \frac{\pi(1-2\sqrt{2}\spac)^2}{\spac^2} \right| \leq \frac{C}{\spac^{2/3}}
    	\eeq
    	(see for instance Theorem 7.7.16 in \cite{H})
so that
    	\beq
    	    \frac{i\Omega_0}{\varepsilon} \int_{\Lambda} d\vec{r} \: g_{\varepsilon}^* \left( \vec{r} \times \nabla g_{\varepsilon} \right) \leq - \frac{\pi^2 \Omega_0}{2 \varepsilon \spac^2} + \frac{C_{\Omega_0,\delta}}{\varepsilon} \leq - \frac{\pi^2 \Omega_0}{2 \delta^2 \varepsilon^2} + \frac{C_{\Omega_0,\delta}}{\varepsilon}.
    	\eeq
    	The first term in \eqref{decomp} is the most difficult to estimate and we deal with it in the following Lemma \ref{kinetic}.
    	\newline
    	Altogether the three upper bounds then give the result for a possibly different constant \( C_{\Omega_0,\delta} \).
\end{proof}

\begin{lem}[Kinetic Energy of Vortices]
    	\label{kinetic}
	\mbox{}	\\
    	Let \( g_{\varepsilon} \) be the function defined in \eqref{phase}, \( \spac = \delta \sqrt{\varepsilon} \) and \( \eta > \frac{5}{2} \). There exists a constant \( C_{\delta} \) independent of \( \varepsilon \) such that for \( \varepsilon \) sufficiently small
    	\beq
        	\int_{\Lambda} d\vec{r} \: \left| \nabla g_{\varepsilon} \right|^2 \leq \frac{\pi^3}{2 \delta^4 \varepsilon^2} + \frac{C_{\delta} |\log \varepsilon|}{\varepsilon}.
    	\eeq
\end{lem}

\begin{proof}
    	We first notice the useful fact that
    	\bdm
        	\int_{\Lambda} d\vec{r} \: \left| \nabla g_{\varepsilon} \right|^2 = \int_{\Lambda} d\vec{r} \: \left| \nabla \phi \right|^2 = \int_{\Lambda} d\vec{r} \: \left| \nabla \tilde{\phi} \right|^2
    	\edm
    	where \( \phi \) is defined in \eqref{phase1} and \( \tilde{\phi} \) is the function
    	\beq
    	\label{phase2}
        	\tilde{\phi}(\vec{r}) =  \sum_{i \in \latt} \ln |\vec{r} - \vec{r}_i|.
    	\eeq
	Indeed,  $\tilde  \phi $ and $ {\phi} $ are conjugate harmonic functions (the real and imaginary parts of $\ln \prod_i(z-z_i)$)) so that $ \partial_x\tilde\phi=-\partial_y\phi$, $\partial_y\tilde \phi=\partial_x\phi$.
	\newline
    	Since $\tilde \phi$  is harmonic, the last integral can be explicitly evaluated by means of partial integration:
    	\bdm
        	\int_{\Lambda} d\vec{r} \: \left| \nabla \tilde{\phi} \right|^2 = \int_{\partial \bo} d\vec{s} \: \cdot \: \frac{\partial \tilde{\phi}}{\partial \vec{n}} \: \tilde{\phi} - \sum_{i \in \latt} \int_{\partial \bi} d\vec{s} \: \cdot \: \frac{\partial \tilde{\phi}}{\partial \vec{n}} \: \tilde{\phi}
    	\edm
    	where \( \vec{n} \) stands for the outer normal to integration path.
    	\newline
    	We are going to consider the two terms separately.

    	\emph{Outer boundary:} The contribution at the outer boundary is given by
   	\beq
	\label{outer}
        	\int_{\partial \bo} d\vec{s} \: \cdot \: \frac{\partial \tilde{\phi}}{\partial \vec{n}} \: \tilde{\phi} = \frac{1}{2} \sum_{i,j \in \latt} \int_0^{2\pi} d\vartheta \: \frac{2-z_j e^{-i\vartheta} - z_j^* e^{i\vartheta}}{|e^{i\vartheta} - z_j|^2} \: \ln \left| e^{i\vartheta} - z_i \right|
    	\eeq
    	where we have used the complex coordinate notation, \( z = x+iy \).
    	\newline
    	The first step in the proof is the replacement of the sum over \( i \) with an integral over \( \mathcal{B}_{1-\frac{3\spac}{\sqrt{2}}} \):
	\bdm
		\sum_{i \in \latt} \ln \left| e^{i\vartheta} - z_i \right|^2 - \frac{1}{\spac^2} \int_{\cup_{i \in \latt} \celli} dz \: \ln \left| e^{i\vartheta} - z \right|^2 = - \frac{1}{\spac^2} \sum_{i \in \latt} \int_{\cell} dz \: \ln \frac{\left| e^{i\vartheta} - z_i - z \right|^2}{\left| e^{i\vartheta} - z_i \right|^2}.
	\edm 
	Thanks to the choice of the lattice \eqref{lattice},
	\bdm
		\frac{\left| e^{i\vartheta} - z_i - z \right|}{\left| e^{i\vartheta} - z_i \right|} > \frac{1}{2}
	\edm
	because, for any \( \vartheta \in [0,2\pi] \), \( \left| e^{i\vartheta} - z_i \right| \geq 2\sqrt{2}\spac \) and \( |z| \leq \spac/\sqrt{2} \). Using therefore the bound
	\bdm
		\ln(1+t) \geq t - t^2
	\edm
	which holds true for any \( t > -1/2 \), we get
	\bdm
		- \frac{1}{\spac^2} \sum_{i \in \latt} \int_{\cell} dz \: \ln \frac{1 - \left| e^{i\vartheta} - z_i - z \right|^2}{\left| e^{i\vartheta} - z_i \right|^2} \leq - \frac{1}{\spac^2} \sum_{i \in \latt} \int_{\cell} d\vec{r} \left\{ \frac{-2\vec{r} \cdot \left( (\cos \vartheta, \sin \vartheta) - \vec{r}_i \right) + r^2}{\left| (\cos \vartheta, \sin\vartheta) - \vec{r}_i \right|^2} + \right.
	\edm
	\bdm
		- \left. \frac{ \left[ -2\vec{r} \cdot \left( (\cos \vartheta, \sin \vartheta) - \vec{r}_i \right) + r^2 \right]^2}{\left| (\cos \vartheta, \sin\vartheta) - \vec{r}_i \right|^4} \right \} \leq
	\edm
	\bdm
		\leq \frac{1}{\spac^2} \sum_{i \in \latt} \int_{\cell} d\vec{r} \: \frac{ 4 \left[ \vec{r} \cdot \left( (\cos \vartheta, \sin \vartheta) - \vec{r}_i \right) \right]^2 + r^4}{\left| (\cos \vartheta, \sin\vartheta) - \vec{r}_i \right|^4} \leq
	\edm
	\bdm
		\leq \frac{1}{\spac^2} \sum_{i \in \latt} \int_{\cell} d\vec{r} \: \left\{ \frac{ 4 r^2}{\left| (\cos \vartheta, \sin\vartheta) - \vec{r}_i \right|^2} + \frac{r^4}{\left| (\cos \vartheta, \sin\vartheta) - \vec{r}_i \right|^4} \right\} \leq  
	\edm
	\bdm
		\leq \sum_{i \in \latt} \left\{ \frac{ C_1 \spac^2}{\left| (\cos \vartheta, \sin\vartheta) - \vec{r}_i \right|^2} + \frac{C_2 \spac^4}{\left| (\cos \vartheta, \sin\vartheta) - \vec{r}_i \right|^4} \right\}.
	\edm
	Since the functions \( 1/r^2 \) and \( 1/r^4 \) are positive and subharmonic we can easily bound the expression above by
	\bdm
		\sum_{i \in \latt} \left\{ \frac{ C_1 \spac^2}{\left| (\cos \vartheta, \sin\vartheta) - \vec{r}_i \right|^2} + \frac{C_2 \spac^4}{\left| (\cos \vartheta, \sin\vartheta) - \vec{r}_i \right|^4} \right\} \leq
	\edm
	\bdm
		\leq \int_{\cup_{i \in \latt} \celli} d\vec{r} \: \left\{ \frac{ C_1 \spac^2}{\left| (\cos \vartheta, \sin\vartheta) - \vec{r} \right|^2} + \frac{C_2 \spac^4}{\left| (\cos \vartheta, \sin\vartheta) - \vec{r} \right|^4} \right\} \leq
	\edm
	\bdm
		\leq \int_{\mathcal{B}_{1-\frac{3\spac}{\sqrt{2}}}} d\vec{r} \: \left\{ \frac{ C_1 \spac^2}{\left| (\cos \vartheta, \sin\vartheta) - \vec{r} \right|^2} + \frac{C_2 \spac^4}{\left| (\cos \vartheta, \sin\vartheta) - \vec{r} \right|^4} \right\} \leq C
	\edm
	so that
	\bdm
		\sum_{i \in \latt} \ln \left| e^{i\vartheta} - z_i \right| \leq \frac{1}{\spac^2} \int_{\cup_{i \in \latt} \celli} dz \: \ln \left| e^{i\vartheta} - z \right| + C .
	\edm
	On the other hand
	\bdm
		\frac{1}{\spac^2} \int_{\cup_{i \in \latt} \celli} dz \: \ln \left| e^{i\vartheta} - z \right| = \frac{1}{\spac^2} \int_{\mathcal{B}_{1-\frac{3\spac}{\sqrt{2}}}} dz \: \ln \left| e^{i\vartheta} - z \right| - \frac{1}{\spac^2} \int_{\mathcal{B}_{1-\frac{3\spac}{\sqrt{2}}} \setminus \cup_{i \in \latt} \celli} dz \: \ln \left| e^{i\vartheta} - z \right| =
	\edm
	\bdm
		= - \frac{1}{\spac^2} \int_{\mathcal{B}_{1-\frac{3\spac}{\sqrt{2}}} \setminus \cup_{i \in \latt} \celli} dz \: \ln \left| e^{i\vartheta} - z \right| \leq \frac{C |\ln \spac| \left| \mathcal{B}_{1-\frac{3\spac}{\sqrt{2}}} \setminus \cup_{i \in \latt} \celli \right|}{\spac^2} \leq 
	\edm
	\bdm
		\leq \frac{C |\ln \spac| \left[ \pi \left(1- \frac{3\spac}{\sqrt{2}} \right)^2 - N_{\varepsilon} \spac^2 \right]}{\spac^2} \leq \frac{C |\ln \spac|}{\spac}  
	\edm
	where we have used the estimate \eqref{number} for the number of points and the fact that
	\bdm
		\int_{\mathcal{B}_R} dz \: \ln \left| e^{i\vartheta} - z \right| = 0
	\edm
	for any \( 0 < R < 1 \).
	\newline
	Since the function
	\bdm
        	a(z) = \frac{2-z e^{-i\vartheta} - z^* e^{i\vartheta}}{|e^{i\vartheta} - z|^2}
    	\edm
	is positive for any \( \vartheta \in [0,2\pi] \) and \( z \in \mathcal{B}_{1-3\varepsilon/\sqrt{2}} \), the initial expression in \eqref{outer} is bounded from above by
	\bdm
		\int_{\partial \bo} d\vec{s} \: \cdot \: \frac{\partial \tilde{\phi}}{\partial \vec{n}} \: \tilde{\phi} \leq \frac{1}{2} \sum_{j \in \latt} \int_0^{2\pi} d\vartheta \: \frac{2-z_j e^{-i\vartheta} - z_j^* e^{i\vartheta}}{|e^{i\vartheta} - z_j|^2} \: \mathcal{R}_{\varepsilon}(\vartheta)
	\edm
	where
	\bdm
		\mathcal{R}_{\varepsilon}(\vartheta) \equiv - \frac{1}{\spac^2} \int_{\mathcal{B}_{1-\frac{3\spac}{\sqrt{2}}} \setminus \cup_{i \in \latt} \celli} dz \: \ln \left| e^{i\vartheta} - z \right| + C
	\edm
	is easily proved to satisfy the upper bound
	\bdm
		\left| \mathcal{R}_{\varepsilon}(\vartheta) \right| \leq \frac{C | \ln\spac | \left| \mathcal{B}_{1-\frac{3\spac}{\sqrt{2}}} \setminus \cup_{i \in \latt} \celli \right|}{\spac^2} \leq \frac{C |\ln\spac|}{\spac}.
	\edm
	We need now to replace the sum over \( j \) with an integral over \( \mathcal{B}_{1-2\sqrt{2}\spac} \): Since the function \( a(z) \) is harmonic, we can apply the mean value theorem to get
    	\bdm
        	a(z_j) - \frac{1}{\spac^2} \int_{\cell} dz \: a(z_j+z) = \frac{1}{\spac^2} \int_{\cell \setminus \bc} dz \: \left[ a(z_j) - a(z_j+z) \right] \equiv b_{\varepsilon}(z_j).
    	\edm
    	 For any \( j \in \latt \), the right hand side can be easily estimated using Harnack's inequality:
    	\bdm
        	b_{\varepsilon}(z_j) \leq \frac{\left| \cell \setminus \bc \right|}{\spac^2} \left[ a(z_j) - \frac{1-\frac{\sqrt{2}\spac}{2}}{1+\frac{\sqrt{2}\spac}{2}} a(z_j) \right] \leq \frac{C \left| \cell \setminus \bc \right| a(z_j)}{\spac} \leq C \spac a(z_j).
    	\edm
    	In the same way it is possible to show that for \( \varepsilon \) sufficiently small, there exists a possibly different constant \( C \) such that
    	\bdm
        	b_{\varepsilon}(z_j) \geq - C \spac a(z_j)
    	\edm
    	so that
    	\bdm
        	\frac{1}{(1+C \spac)\spac^2} \int_{\cell} dz \: a(z_j+z) \leq a(z_j) \leq \frac{1}{(1-C\spac)\spac^2} \int_{\cell} dz \: a(z_j+z)
    	\edm
    	and then
    	\bdm
        	\left| a(z_j) - \frac{1}{\spac^2} \int_{\cell} dz \: a(z_j+z) \right| \leq C \spac.
    	\edm
    	Since
    	\bdm
        	\sum_{j \in \latt} \int_{\cell} dz \: a(z_j+z) = \int_{\cup_j \cellj} dz \: a(z)
   	\edm
    	and
    	\bdm
        	0 \leq \int_{\mathcal{B}_{1-2\sqrt{2}\spac} \setminus \cup_j \cellj} dz \: a(z) \leq \int^{1-\frac{3\spac}{\sqrt{2}}}_{1-2\sqrt{2}\spac} dr \: r \int_0^{2 \pi} d\gamma \: a(re^{i\gamma}) \leq C \spac
    	\edm
    	we conclude that
    	\beq
    	\label{est1}
    		\left|\mathcal{R}^{\prime}_{\varepsilon}(\vartheta) \right| \equiv \left| \sum_{j \in \latt} \frac{2-z_j e^{-i\vartheta} - z_j^* e^{i\vartheta}}{|e^{i\vartheta} - z_j|^2} - \frac{1}{\spac^2} \int_{\mathcal{B}_{1-2\sqrt{2}\spac}} dz \: \frac{2-z e^{-i\vartheta} - z^* e^{i\vartheta}}{|e^{i\vartheta} - z|^2} \right| \leq C \spac N_{\varepsilon}.
    	\eeq
    	On the other hand, using again the harmonicity of \( a \), one has
    	\bdm
        	\frac{1}{\spac^2} \int_{\mathcal{B}_{1-2\sqrt{2}\spac}} dz \: \frac{2-z e^{-i\vartheta} - z^* e^{i\vartheta}}{|e^{i\vartheta} - z|^2} = \frac{2 \pi \left( 1 - 2\sqrt{2}\spac \right)^2}{\spac^2}.
    	\edm
    	Since for the fundamental solution of the Laplace equation
    	\beq
    	\label{Green}
        	\frac{1}{2\pi} \int_0^{2\pi} d\vartheta \: \ln \left| R e^{i\vartheta} - \vec{x} \right| = \left\{
        	\begin{array}{ll}
            		\ln R       &   \mbox{if} \:\: |\vec{x}| \leq R \\
            		\ln |\vec{x}|   &   \mbox{if} \:\: |\vec{x}| \geq R
        	\end{array}
        	\right.
    	\eeq
    	we then get altogether
    	\bdm
    		\int_{\partial \bo} d\vec{s} \: \cdot \: \frac{\partial \tilde{\phi}}{\partial \vec{n}} \: \tilde{\phi} \leq - \frac{2 \pi \left( 1 - 2\sqrt{2}\spac \right)^2}{\spac^4} \int_{\mathcal{B}_{1-\frac{3\spac}{\sqrt{2}}} \setminus \cup_j \cellj} dz \int_0^{2 \pi} d\vartheta \: \ln \left| e^{i\vartheta} - z \right | + \int_0^{2\pi} d\vartheta \: \left| \mathcal{R}^{\prime}_{\varepsilon}(\vartheta) \right| \left| \mathcal{R}_{\varepsilon}(\vartheta) \right| + \frac{C}{\spac^2} \leq 
    	\edm
	\beq
	\label{est2}
        	\leq 2\pi \sup_{\vartheta \in [0,2\pi]} \left| \mathcal{R}^{\prime}_{\varepsilon}(\vartheta) \right| \left| \mathcal{R}_{\varepsilon}(\vartheta) \right| + \frac{C}{\spac^2} \leq \frac{C |\ln \spac|}{\spac^2} + \frac{C}{\spac^2} \leq \frac{C_{\delta}|\log\varepsilon|}{\varepsilon}.
	\eeq	

    	\emph{Inner boundary:} A straightforward calculation gives
    	\bdm
        	- \sum_{i \in \latt} \int_{\partial \bi} d\vec{s} \: \cdot \: \frac{\partial \tilde{\phi}}{\partial \vec{n}} \: \tilde{\phi} = 2 \pi \eta N_{\varepsilon} | \ln \varepsilon | - \sum_{\substack{i,j \in \latt \\ i \neq j }} \int_0^{2 \pi} d\vartheta \: \ln \left| \varepsilon^{\eta}e^{i\vartheta} + z_i - z_j \right| +
    	\edm
    	\bdm
        	- \frac{\varepsilon^{\eta}}{2} \sum_{\substack{i,j \in \latt \\ i \neq j }} \int_0^{2\pi} d\vartheta \: \tilde{\phi}(z_i + \varepsilon^{\eta} e^{i\vartheta}) \: \frac{\varepsilon^{\eta} - (z_i-z_j)e^{-i\vartheta} - (z_i-z_j)^*e^{i\vartheta}}{\left| \varepsilon^{\eta}e^{i\vartheta} + z_i - z_j \right|^2} \leq
    	\edm
    	\bdm
        	\leq 2 \pi \eta N_{\varepsilon} | \ln \varepsilon | - 2 \pi \sum_{\substack{i,j \in \latt \\ i \neq j }} \ln \left| z_i - z_j \right| + \frac{C \varepsilon^{\eta} N_{\varepsilon}}{\spac^2} \sum_{i \in \latt} \int_0^{2\pi} d\vartheta \: | \tilde{\phi}(z_i + \varepsilon^{\eta} e^{i\vartheta}) | \leq
    	\edm
    	\begin{equation}
        	\leq 2 \pi \eta N_{\varepsilon} | \ln \varepsilon | - 2 \pi \sum_{\substack{i,j \in \latt \\ i \neq j }} \ln \left| z_i - z_j \right| + \frac{C \varepsilon^{\eta} N^2_{\varepsilon} |\log \varepsilon|}{\spac^2} \leq
    	 \frac{C_{\delta} |\log\varepsilon|}{\varepsilon} - 2 \pi \sum_{\substack{i,j \in \latt \\ i \neq j }} \ln \left| z_i - z_j \right|\label{innerbd}
    	\end{equation}
    	where we have used \eqref{nabla}.
    	\newline
    	In order to get the desired estimate we need now to replace the sum over one of the two indices in the expression above with the integration on a suitable domain. Therefore the quantity which has to be estimated is the difference
	\beq
	\label{logaritm}
        	\sum_{\substack{i,j \in \latt \\ i \neq j }} \left\{ - \ln \left| \vec{r}_i - \vec{r}_j \right| + \frac{1}{\spac^4} \int_{\cell} d\vec{r} \int_{\cell} d\vec{r}^{\prime} \: \ln \left| \vec{r}_i - \vec{r}_j + \vec{r} - \vec{r}^{\prime} \right| \right\}. 
    	\eeq
	Using the estimate \( \ln(t) \leq \frac{1}{2} (t^2-1) \), which holds for any \( t > 0 \), we can bound the expression under the sum in the following way
	\bdm
		\frac{1}{\spac^4} \sum_{\substack{i,j \in \latt \\ i \neq j }} \int_{\cell} d\vec{r} \int_{\cell} d\vec{r}^{\prime} \: \ln \frac{\left| \vec{r}_i - \vec{r}_j + \vec{r} - \vec{r}^{\prime} \right|}{\left| \vec{r}_i - \vec{r}_j \right|} \leq \frac{1}{2\spac^4} \sum_{\substack{i,j \in \latt \\ i \neq j }} \int_{\cell} d\vec{r} \int_{\cell} d\vec{r}^{\prime} \: \left\{ \frac{\left| \vec{r}_i - \vec{r}_j + \vec{r} - \vec{r}^{\prime} \right|^2}{\left| \vec{r}_i - \vec{r}_j \right|^2} - 1 \right\} =
	\edm
	\bdm
		= \sum_{\substack{i,j \in \latt \\ i \neq j }} \frac{1}{2\spac^4 \left| \vec{r}_i - \vec{r}_j \right|^2} \int_{\cell} d\vec{r} \int_{\cell} d\vec{r}^{\prime} \: \left[ \left| \vec{r} - \vec{r}^{\prime} \right|^2 + 2 \left( \vec{r} - \vec{r}^{\prime} \right) \cdot \left( \vec{r}_i - \vec{r}_j \right) \right] =
	\edm
	\bdm
		= \sum_{\substack{i,j \in \latt \\ i \neq j }} \frac{1}{2\spac^4 \left| \vec{r}_i - \vec{r}_j \right|^2} \int_{\cell} d\vec{r} \int_{\cell} d\vec{r}^{\prime} \: \left( r^2 + {r^{\prime}}^2 \right) \leq \sum_{\substack{i,j \in \latt \\ i \neq j }} \frac{C \spac^2}{\left| \vec{r}_i - \vec{r}_j \right|^2}
	\edm
	where we have used the central symmetry of the fundamental cell \( \cell \) and the lattice \( \latt \). 
	\newline
	On the other hand, since the function \( 1/r^2 \) is subharmonic and positive, one can easily prove that
	\bdm
        	\sum_{\substack{i,j \in \latt \\ i \neq j }} \frac{C \spac^2}{\left| \vec{r}_i - \vec{r}_j \right|^2} \leq \frac{C}{\spac^2} \sum_{\substack{i,j \in \latt \\ i \neq j }} \int_{\cell} d\vec{r} \int_{\cell} d\vec{r}^{\prime} \: \frac{1}{\left| \vec{r}_i - \vec{r}_j + \vec{r} - \vec{r}^{\prime} \right|^2} \leq 
	\edm
	\bdm
		\leq \frac{C}{\spac^2} \sum_{\substack{i,j \in \latt \\ i \neq j }} \int_{\celli} d\vec{r} \int_{\cellj} d\vec{r}^{\prime} \: \frac{1}{\left| \vec{r} - \vec{r}^{\prime} \right|^2} \leq \frac{C}{\spac^2} \int_{\frac{\spac}{2}}^2 \frac{dr}{r} \leq \frac{C_{\delta} |\log\varepsilon|}{\varepsilon}
    	\edm
	so that the difference in \eqref{logaritm} is bounded by \( C_{\delta} |\log\varepsilon|/\varepsilon \).
	\newline
    	In order to extend the integration to the whole disc \( \bo \), we observe that
    	\bdm
        	- \frac{1}{\spac^4} \sum_{\substack{i,j \in \latt \\ i \neq j }} \int_{\celli} d\vec{r} \int_{\cellj} d\vec{r}^{\prime} \: \ln \left| \vec{r} - \vec{r}^{\prime} \right| \leq - \frac{1}{\spac^4} \int_{\cup_{i \in \latt} \celli} d\vec{r} \int_{\cup_{j \in \latt} \cellj} d\vec{r}^{\prime} \: \ln \left| \vec{r} - \vec{r}^{\prime} \right| =
    	\edm
	\beq
	\label{est3}
		= - \frac{1}{\spac^4} \int_{\bo} d\vec{r} \int_{\cup_{i \in \latt} \celli} d\vec{r}^{\prime} \: \ln \left| \vec{r} - \vec{r}^{\prime} \right| + \frac{1}{\spac^4} \int_{\mathcal{A}_{\varepsilon}} d\vec{r} \int_{\bo} d\vec{r}^{\prime} \: \ln \left| \vec{r} - \vec{r}^{\prime} \right| - \frac{1}{\spac^4} \int_{\mathcal{A}_{\varepsilon}} d\vec{r} \int_{\mathcal{A}_{\varepsilon}} d\vec{r}^{\prime} \: \ln \left| \vec{r} - \vec{r}^{\prime} \right|
	\eeq
    	where \( \mathcal{A}_{\varepsilon} \) stands for the domain \( \bo \setminus \cup_{i \in \latt} \celli \).
	\newline
	The last term in the expression above is bounded by
	\bdm
		- \frac{1}{\spac^4} \int_{\mathcal{A}_{\varepsilon}} d\vec{r} \int_{\mathcal{A}_{\varepsilon}} d\vec{r}^{\prime} \: \ln \left| \vec{r} - \vec{r}^{\prime} \right| \leq - \frac{1}{\spac^4} \int_{\tilde{\mathcal{A}}_{\varepsilon}} d\vec{r} \int_{\tilde{\mathcal{A}}_{\varepsilon}} d\vec{r}^{\prime} \: \ln \left| r - r^{\prime} \right| \leq - \frac{C \ln \spac}{\spac^2} \leq \frac{C_{\delta} | \log \varepsilon |}{\varepsilon}
	\edm
	where \( \tilde{\mathcal{A}}_{\varepsilon} \equiv \bo \setminus \mathcal{B}_{1 - 2\sqrt{2} \spac} \).
	\newline 
	For the second term in \eqref{est3}, we can use \eqref{Green} to get
	\bdm
		\frac{1}{\spac^4} \int_{\mathcal{A}_{\varepsilon}} d\vec{r} \int_{\bo} d\vec{r}^{\prime} \: \ln \left| \vec{r} - \vec{r}^{\prime} \right| = \frac{2 \pi}{\spac^4} \int_{\mathcal{A}_{\varepsilon}} d\vec{r} \left\{ \int_0^r dr^{\prime} r^{\prime} \: \ln r + \int_r^1 dr^{\prime} r^{\prime} \: \ln r^{\prime} \right\} \leq 0.
	\edm
	Therefore one has from \eqref{est3} 
	\bdm
		- \frac{1}{\spac^4} \sum_{\substack{i,j \in \latt \\ i \neq j }} \int_{\celli} d\vec{r} \int_{\cellj} d\vec{r}^{\prime} \: \ln \left| \vec{r} - \vec{r}^{\prime} \right| \leq - \frac{1}{\spac^4} \int_{\bo} d\vec{r} \int_{\cup_{i \in \latt} \celli} d\vec{r}^{\prime} \: \ln \left| \vec{r} - \vec{r}^{\prime} \right| + \frac{C_{\delta}| \log \varepsilon |}{\varepsilon} \leq 
	\edm
	\bdm
		\leq \frac{\pi}{2 \spac^4} \int_{\bo} d\vec{r}^{\prime} \: \left( 1 - {r^{\prime}}^2 \right) + \frac{C_{\delta}| \log \varepsilon |}{\varepsilon} \leq \frac{\pi^2}{4 \spac^4} + \frac{C_{\delta}| \log \varepsilon |}{\varepsilon}
	\edm
    	so that
	\bdm
		- 2 \pi \sum_{\substack{i,j \in \latt \\ i \neq j }} \ln \left| z_i - z_j \right| \leq \frac{\pi^3}{2 \spac^4} + \frac{C_{\delta}| \log \varepsilon |}{\varepsilon}
	\edm
	and finally
    	\beq
        	- \sum_{i \in \latt} \int_{\partial \bi} d\vec{s} \: \cdot \: \frac{\partial \tilde{\phi}}{\partial \vec{n}} \: \tilde{\phi} \leq  \frac{\pi^3}{2 \spac^4} + \frac{C_{\delta} |\log\varepsilon|}{\varepsilon}.
    	\eeq
    	
	Combining this result with the estimate for the contribution at the outer boundary, we complete the proof.
\end{proof}

\subsection{The Regime $\Omega(\varepsilon) \gg 1/\varepsilon$}

\begin{proof1}{Theorem \ref{energya}}
	The lower bound can be proved in the same way as in the proof of Theorem \ref{energy}, so that one easily gets
	\bdm
		\gpe \geq \frac{\tfea}{\varepsilon^{2+2\alpha}}.
	\edm
	For the upper bound we evaluate the functional on the following trial function
	\beq
		\label{triala}
        	\trial(\vec{r}) = \tilde{c}_{\varepsilon} j_{\varepsilon}(r) \: \sqrt{\tfma(r)} \: \exp \left\{ i \left[ \frac{\Omega_1}{2 \varepsilon^{1+\alpha}} \right] \vartheta \right\}
	\eeq
	where we used polar coordinates, \( \vec{r} = (r ,\vartheta) \), \( [ \:\: \cdot \:\: ] \) stands for the integer part, \( j_{\varepsilon} \) is the cut-off function
	\beq
		\label{cutprofa}
        	j_{\varepsilon}(r) = \left\{
        	\begin{array}{ll}
              		0   &   \mbox{if} \:\:\:\: r \leq \rtfa \\
              		\mbox{} &   \mbox{} \\
              		\displaystyle{\frac{r^2-\rtfa^2}{\varepsilon^{\beta}}}  & \mbox{if} \:\:\:\: \rtfa^2 \leq r^2 \leq \rtfa^2 + \varepsilon^{\beta}  \\
              		\mbox{} &   \mbox{} \\
              		1   &   \mbox{otherwise}
        	\end{array}
        	\right.
	\eeq
	with some \( \beta > \alpha \), and \( \tilde{c}_{\varepsilon} \) is a normalization constant satisfying the following bounds
	\beq
		\label{coeffnorm}
		1 < \tilde{c}_{\varepsilon}^2 \leq 1 + C \varepsilon^{2\beta-2\alpha}.
	\eeq
	A simple calculation shows that
	\bdm
		\gpf[\trial] = \tilde{c}_{\varepsilon}^2 \int_{\bo} d\vec{r} \: \left| \partial_r \left( j_{\varepsilon} \sqrt{\tfma} \right) \right|^2 + \tilde{c}_{\varepsilon}^2 \int_{\bo} d\vec{r} \: j_{\varepsilon}^2 \: \tfma \: \left\{ \frac{1}{r} \left[ \frac{\Omega_1}{2 \varepsilon^{1+\alpha}} \right] - \frac{\Omega_1 r}{2 \varepsilon^{1+\alpha}} \right\}^2 +
	\edm
	\beq
		\label{upperbounda}
        	+ \frac{\tff[\tilde{c}_{\varepsilon}^2 j_{\varepsilon}^2 \tfma]}{\varepsilon^{2+2\alpha}}.
	\eeq
	The first term in \eqref{upperbounda} is bounded by (using \eqref{coeffnorm})
	\bdm
		\tilde{c}_{\varepsilon}^2 \int_{\bo} d\vec{r} \: \left| \partial_r \left( j_{\varepsilon} \sqrt{\tfma} \right) \right|^2 \leq 2\tilde{c}_{\varepsilon}^2 \int_{\bo} d\vec{r} \: \left| \partial_r j_{\varepsilon} \right|^2 \tfma + \tilde{c}_{\varepsilon}^2 \int_{\bo} d\vec{r} \: \frac{j_{\varepsilon}^2}{2 \tfma} \: \left( \frac{\partial \tfma}{\partial r} \right)^2 \leq
	\edm
	\bdm
		\leq \frac{C_1}{\varepsilon^{2 \beta}} \int_{\rtfa}^{\sqrt{\rtfa^2+\varepsilon^{\beta}}} dr r ^3\: \tfma(r) + \frac{C_2}{\varepsilon^{\beta-2\alpha}} \int_{\rtfa}^{\sqrt{\rtfa^2 + \varepsilon^{\beta}}} dr r (r^2-R_\varepsilon^2)\left( \frac{\partial \tfma}{\partial r} \right)^2 +
	\edm
	\bdm
		+ C_3 \varepsilon^{2\alpha} \int_{\sqrt{\rtfa^2+\varepsilon^{\beta}}}^1 dr \frac{r}{r^2 - \rtfa^2} \: \left( \frac{\partial \tfma}{\partial r} \right)^2 \leq
	\edm
     	\bdm
        	\leq \frac{C_1}{\varepsilon^{2\alpha}}+ \frac{C_2}{\varepsilon^{2\alpha}} + \frac{C_3 |\log \varepsilon|}{\varepsilon^{2\alpha}} \leq \frac{C_{\Omega_1} |\log \varepsilon|}{\varepsilon^{2\alpha}}.
	\edm
     	(The constants depend on the choice of $\beta$ and $C_3\to\infty$ if $\beta\to\infty$.) Moreover, using \eqref{coeffnorm} and the fact that
     	\bdm
        	\left[ \frac{\Omega_1}{2 \varepsilon^{1+\alpha}}  \right] = \frac{\Omega_1}{2 \varepsilon^{1+\alpha}} - \kappa_{\varepsilon}
     	\edm
     	for some \( 0 \leq \kappa_{\varepsilon} < 1 \), we can estimate the second term in \eqref{upperbounda} as follows
     	\bdm
        	\tilde{c}_{\varepsilon}^2 \int_{\bo} d\vec{r} \: j_{\varepsilon}^2 \: \tfma \: \left\{ \frac{1}{r} \left[ \frac{\Omega_1}{2 \varepsilon^{1+\alpha}} \right] - \frac{\Omega_1 r}{2 \varepsilon^{1+\alpha}} \right\}^2 \leq \frac{\tilde{c}_{\varepsilon}^2\pi \Omega_1^2}{2 \varepsilon^{2+2\alpha}} \int_{\rtfa}^{1} dr r \: \tfma \: \left( \frac{1}{r} - r \right)^2 +
     	\edm
     	\bdm
        	+ \tilde{c}_{\varepsilon}^2 \kappa_{\varepsilon}^2 2\pi\int_{\rtfa}^1 dr \: \frac{\tfma}{r} \leq \frac{C_1 \left(1 - \rtfa^2 \right)^2}{\varepsilon^{2+2\alpha}} + C_2 \leq \frac{C_{\Omega_1}}{\varepsilon^2}.
     	\edm
     	In a similar way one can prove that
     	\bdm
        	\tffa[\tilde{c}_{\varepsilon}^2 j_{\varepsilon}^2 \tfma] \leq \tffa[\tfma] + 2 \pi \left( \tilde{c}_{\varepsilon}^4 - 1 \right) \int_{\rtfa}^1 dr \: r {\tfma}^2(r) + \frac{\pi \Omega_1^2}{2} \int_{\rtfa}^{\sqrt{\rtfa^2+\varepsilon^{\beta}}} dr \: r^3 \left( 1-j_{\varepsilon}^2 \right) \tfma(r) \leq
     	\edm
 	\bdm
  		\leq \tfea + C_1 \varepsilon^{2\beta-4\alpha} \int_{\rtfa^2}^1 dz \: \left( z - \rtfa^2 \right)^2 + \frac{C_2}{\varepsilon^{2\alpha}} \int_{\rtfa^2}^{\rtfa^2+\varepsilon^{\beta}} dz \: z \left[ 1- \left( \frac{z-\rtfa^2}{\varepsilon^{\beta}} \right)^2 \right] \left( z-\rtfa^2 \right) \leq
 	\edm
 	\bdm
  		\leq \tfea + C_1 \varepsilon^{2\beta-\alpha} + \frac{C_2}{\varepsilon^{2\alpha}} \int_{0}^{\varepsilon^{\beta}} dz \: z \left( 1-\frac{z^2}{\varepsilon^{2\beta}} \right) \leq \tfea + C_{\Omega_1} \varepsilon^{2\beta - 2\alpha}
 	\edm
     	and then the result follows if we choose a finite \(\beta > 2\alpha \).
\end{proof1}

\begin{proof1}{Corollary \ref{profilea}}
    	Let us start by considering the case \( 0 < \alpha < 2 \). Defining
	$\dtfa={\mathcal B}_1\setminus {\mathcal B}_{R_\varepsilon}$
	we first notice that for any non negative function \( \rho \in L^2(\dtfa) \), normalized to 1 in \( L^1(\dtfa) \),
    	\bdm
        	\tffa\left[\rho,\dtfa \right] \geq \tfea
    	\edm
    	where \( \tffa[\rho,\mathcal{D_\varepsilon}] \) denotes the functional
    	\bdm
        	\tffa[\rho,\mathcal{D_\varepsilon}] \equiv \varepsilon^{2\alpha} \int_{\mathcal{D_\varepsilon}} d\vec{r} \: \left\{ \rho^2 - \frac{\Omega_1^2 r^2 \rho}{4 \varepsilon^{2\alpha}} \right\}.
    	\edm
    	Hence, setting \( \rho_{\varepsilon} \equiv | \gpm |^2 \) and
    	\bdm
        	\tilde{\rho}_{\varepsilon} \equiv \frac{\rho_{\varepsilon}}{\left\| \rho_{\varepsilon} \right\|_{L^1(\dtfa)}}
    	\edm
    	we get
    	\bdm
        	\tffa \left[ \rho_{\varepsilon} \right] = \left\|  \rho_{\varepsilon} \right\|_{L^1(\dtfa)} \tffa\left[ \tilde{\rho}_{\varepsilon},\dtfa \right] + \tffa\left[\rho_{\varepsilon} , \bo \setminus \dtfa \right] +
    	\edm
    	\bdm
        	+ \varepsilon^{2\alpha} \left\| \rho_{\varepsilon} \right\|^2_{L^2(\dtfa)} \left( 1 - \frac{1}{\left\| \rho_{\varepsilon} \right\|_{L^1(\dtfa)}} \right) \geq
    	\edm
    	\bdm
        	\geq \tfea \left\|  \rho_{\varepsilon} \right\|_{L^1(\dtfa)} + \tffa\left[\rho_{\varepsilon} , \bo \setminus \dtfa \right] + \varepsilon^{2\alpha} \left\| \rho_{\varepsilon} \right\|^2_{L^2(\dtfa)} \left( 1 - \frac{1}{\left\| \rho_{\varepsilon} \right\|_{L^1(\dtfa)}} \right) \geq
    	\edm
    	\bdm
        	\geq \tfea \left\|  \rho_{\varepsilon} \right\|_{L^1(\dtfa)} - \frac{\Omega_1^2 \rtfa^2}{4} \left( 1 - \left\| \rho_{\varepsilon} \right\|_{L^1(\dtfa)} \right) + \varepsilon^{2\alpha} \left\| \rho_{\varepsilon} \right\|^2_{L^2(\dtfa)} \left( 1 - \frac{1}{\left\| \rho_{\varepsilon} \right\|_{L^1(\dtfa)}} \right) \geq
    	\edm
    	\bdm
        	\geq \tfea \left\|  \rho_{\varepsilon} \right\|_{L^1(\dtfa)} - \frac{\Omega_1^2 \rtfa^2}{4} \left( 1 - \left\| \rho_{\varepsilon} \right\|_{L^1(\dtfa)} \right) + \varepsilon^{2\alpha} \left( \left\|  \rho_{\varepsilon} \right\|_{L^1(\dtfa)} - 1 \right) \frac{\left\|  \rho_{\varepsilon} \right\|_{L^1(\dtfa)}}{\left| \dtfa \right|}
    	\edm
    	where in the last step we have used  Schwarz's inequality and the fact that \( \left\|  \rho_{\varepsilon} \right\|_{L^1(\dtfa)} \leq 1 \).
    	\newline
    	On the other hand from the upper bound in the proof of Theorem \ref{energya} one has
    	\bdm
        	\tffa \left[ \rho_{\varepsilon} \right] \leq  \varepsilon^{2+2\alpha} \gpf \left[ \gpm \right] \leq \tfea + C_1 \varepsilon^{2\alpha} + C_2 \varepsilon^2 |\log \varepsilon|
    	\edm
    	and then (omitting for simplicity the subscript \( L^1(\dtfa) \)),
    	\bdm
        	\tfea \left\|  \rho_{\varepsilon} \right\| - \left[ \frac{\Omega_1^2 \rtfa^2}{4} + \frac{\varepsilon^{2\alpha} \left\| \rho_{\varepsilon} \right\|}{\left| \dtfa  \right|} \right]  \left( 1 - \left\| \rho_{\varepsilon} \right\| \right) \leq \tfea + C_1 \varepsilon^{2\alpha} + C_2 \varepsilon^2 |\log \varepsilon|
    	\edm
  	and therefore
   	\bdm
        	\left[ - \frac{\Omega_1 \varepsilon^{\alpha}}{3 \sqrt{\pi}} + \frac{\Omega_1 \varepsilon^{\alpha}}{4 \sqrt{\pi}} \left\| \rho_{\varepsilon} \right\|  \right]  \left( 1 - \left\| \rho_{\varepsilon} \right\| \right) + C_1 \varepsilon^{2\alpha} + C_2 \varepsilon^2 |\log \varepsilon| \geq 0
    	\edm
    	or
    	\bdm
        	\left\| \rho_{\varepsilon} \right\|^2 - \frac{7}{3} \left\| \rho_{\varepsilon} \right\| + \frac{4}{3} - C_1 \varepsilon^{\alpha} - C_2 \varepsilon^{2-\alpha} |\log \varepsilon| \leq 0
    	\edm
    	which implies that, for \( \varepsilon \) sufficiently small,
    	\bdm
        	\left\|  \rho_{\varepsilon} \right\|_{L^1(\dtfa)} \geq 1 - C_{\Omega_1} \varepsilon^{\alpha} - C_{\Omega_1}^{\prime} \varepsilon^{2-\alpha} |\log \varepsilon|.
    	\edm
    	The result therefore follows from the normalization of \( \rho_{\varepsilon} \) in \( L^1(\bo) \).
    	\newline
    	If \( \alpha \geq 2 \), the upper bound contained in the proof of Theorem \ref{energya} gives immediately the following bound
    	\bdm
        	\int_{\bo} d\vec{r} \: r^2 \: \left| \gpm \right|^2 \geq 1 - C \varepsilon^2 |\log \varepsilon|
    	\edm
    	and then, using the normalization of \( \gpm \), the result is proved.
\end{proof1}

\begin{proof1}{Proposition \ref{expa}}
    	Let us first consider the case \( 0 < \alpha < 2 \): as in the proof of Eq.\  \eqref{pointwise}, we first need to prove a pointwise estimate for \( U_{\varepsilon} \equiv \left| \gpm \right|^2 \) and, in order to find such an upper bound, we have to estimate the chemical potential, appearing in the variational equation satisfied by \( \gpm \). From the definition of the chemical potential and the upper bound contained in the proof of Theorem \ref{energya}, we immediately get
    	\bdm
        	\varepsilon^{2+2\alpha} \chem \leq \tfea + C_1 \varepsilon^{2\alpha} + C_2 \varepsilon^{2} |\log \varepsilon| + \varepsilon^{2\alpha} \left\| \gpm \right\|^4_{L^4(\bo)}
    	\edm
    	but, from the same upper bound we obtain
    	\bdm
        	\varepsilon^{2 \alpha} \left\| \gpm \right\|^4_{L^4(\bo)} \leq \frac{\Omega_1^2}{4} \int_{\bo} d\vec{r} \: r^2 \left| \gpm \right|^2 + \tfe + C_1 \varepsilon^{2 \alpha} + C_2 \varepsilon^{2} | \log \varepsilon | \leq
    	\edm
    	\bdm
        	\leq \frac{\Omega_1^2}{4} + \tfe + C_1 \varepsilon^{2 \alpha} + C_2 \varepsilon^{2} | \log \varepsilon |
    	\edm
    	and then
    	\bdm
        	\varepsilon^{2+2\alpha} \chem \leq 2 \tfe + \frac{\Omega_1^2}{4} + C_1 \varepsilon^{2 \alpha} + C_2 \varepsilon^{2} | \log \varepsilon | \leq
    	\edm
    	\beq
    	\label{chema}
        	\leq - \frac{\Omega_1^2}{4} + \frac{4 \Omega_1 \varepsilon^{\alpha}}{3 \sqrt{\pi}} + C_1 \varepsilon^{2 \alpha} + C_2 \varepsilon^{2} | \log \varepsilon |.
    	\eeq
    	By replacing the above bound in the variational equation, we get
    	\bdm
        	-\frac{1}{2} \Delta U_{\varepsilon} \leq \frac{\Omega_1^2}{4} \left[ r^2 - 1 + \frac{16 \varepsilon^{\alpha}}{3 \sqrt{\pi} \Omega_1} + C_1 \varepsilon^{2 \alpha} + C_2 \varepsilon^{2} | \log \varepsilon | - C_3 \varepsilon^{2 \alpha} U_{\varepsilon}  \right] \frac{U_{\varepsilon}}{\varepsilon^{2+2\alpha}}
    	\edm
    	and we can conclude that there exists a constant \( c \) (depending on \( \Omega_1 \)), such that the function \( U_{\varepsilon}(\vec{r}) \) is subharmonic for any \( r^2 \leq 1 - c \varepsilon^{\alpha} \). Following again the proof of Eq.\  \eqref{pointwise}, we can therefore obtain the following estimate
    	\bdm
        	U_{\varepsilon}(\vec{r}) \leq \frac{C \| U_{\varepsilon} \|_{L^1(\bo \setminus \dtfa)}}{\varrho^2}
    	\edm
    	for any \( \vec{r} \in \bo \)  such that
    	\bdm
        	r \leq \sqrt{1 - c\varepsilon^{\alpha}} - \varrho.
    	\edm
    	Choosing for instance \( \varrho = \varepsilon^{\alpha^{\prime}}/3 \) and using \eqref{l2norma1}, we can conclude that there exists a constant \( C_{\Omega_1} \) such that
    	\bdm
        	U_{\varepsilon}(\vec{r}) \leq C_{\Omega_1} \varepsilon^{\frac{\alpha^{\prime}}{3}} |\log \varepsilon|
    	\edm
    	for any \( \vec{r} \in \detf '\). The result then follows from the application of the comparison principle to the variational equation satisfied in \( \detf '\) by
    	\bdm
    	    	U_{\varepsilon}^{\prime} \equiv \frac{U_{\varepsilon}}{C_{\Omega_1} \varepsilon^{\frac{\alpha^{\prime}}{3}} |\log \varepsilon|}.
   	\edm
    	The case \( \alpha \geq 2 \) can be treated in a similar way. The only difference is in the upper bound for the chemical potential \eqref{chema}, which in this case becomes
    	\bdm
        	\varepsilon^{2+2\alpha} \chem \leq - \frac{\Omega_1^2}{4} + C \varepsilon^2 |\log \varepsilon|.
    	\edm
    	The subharmonicity of \( U_{\varepsilon} \) can now be proved in the region \( r \leq 1 - \varepsilon^{\beta} \), for any \( 1 \leq \beta < 2 \). As a straightforward consequence of \eqref{l2norma2}, we then get the pointwise estimate
    	\bdm
        	U_{\varepsilon}(\vec{r}) \leq C_{\Omega_1} \varepsilon^{\frac{2-\beta}{3}} |\log \varepsilon|
    	\edm
    	for any \( \vec{r} \in \detf^{\prime\prime} \). The result is again obtained by means of the comparison principle.
\end{proof1}

\section{Conclusions and Perspectives}

We have analyzed rigorously the leading order asymptotics for the ground state energy and the density profile of a rapidly rotating Bose-Einstein condensate in a flat  trap with a finite radius in the limit where the coupling parameter is large. Depending on the scaling of the rotational velocity with the coupling parameter, different asymptotic density functionals emerge. 

Our estimates are based on trial functions that capture the essential features of the expected vortex structure and show the possible formation of ``holes'' where the density is exponentially small as a function of the inverse coupling parameter. The error terms in our estimates are of the expected order but the bounds are not sharp enough to exhibit the details of the fine vortex structure. Nevertheless,
we can prove that rotational symmetry is broken in the ground state for ${\rm const.}|\log\varepsilon|<\Omega(\varepsilon)\lesssim \mathrm{const.}/\varepsilon$.

An important open problem is to carry the analysis further to the next to leading order and investigate the transition of the vortex lattice to a ``giant vortex'' at high rotational velocities.

\vspace{1cm}
\emph{Acknowledgments:}  M.C. is grateful to S. Fournais and B. Helffer for helpful comments and suggestions. We  thank R. Seiringer for pointing out a mistake in a previous version of Section 2.1. This work is supported by the EU Post Doctoral Training Network HPRN-CT-2002-00277 ``Analysis and Quantum'' and the Austrian Science Fund (FWF) grant P17176-N02.

\end{document}